\documentclass[superscriptaddress,aps,pra,amssymb,amsfonts,letterpaper]{revtex4}
\usepackage[margin=0.5in]{geometry}
\usepackage{hyperref}
\usepackage{enumerate}
\usepackage{verbatim}
\usepackage{amsmath}
\usepackage{latexsym}
\usepackage{revsymb}
\usepackage{yfonts}
\usepackage{amsfonts}
\usepackage{amsmath}
\usepackage{amssymb}
\usepackage{amsthm}
\usepackage{graphicx}
\usepackage{bm}
\usepackage{bbm}
\usepackage{color}
\usepackage{mathrsfs}

\def\pij{p_{i\to j}}
\def\dmin{D_{\min}}
\def\dmax{D_{\max}}
\def\<{\langle}
\def\>{\rangle}
\def\ot{\otimes}
\def\ermax{E_R^{\max}}
\def\esmax{E_S^{\max}}
\def\ecal{{\cal E}}
\def\hcal{{\cal H}}
\def\ecalr{{\ecal_R}}
\def\etaE{\eta_{E-E_S}^R}

\def\ideta{\eta}
\def\rhor{\rho_R}
\def\rhos{\rho_S}

\def\rk{{\rm rank}}

\newcommand{\be}{\begin{eqnarray} \begin{aligned}}
\newcommand{\ee}{\end{aligned} \end{eqnarray} }
\newcommand{\benn}{\begin{eqnarray*} \begin{aligned}}
\newcommand{\eenn}{\end{aligned} \end{eqnarray*} }

\newcommand{\bc}{\begin{center}}
\newcommand{\ec}{\end{center}}

\newcommand{\id}{\mathbb{I}}

\newcommand{\tr}{\mathop{\mathsf{tr}}\nolimits}

\newcommand{\e}{\mathrm{e}}

\newcommand{\bee}{\begin{enumerate}}
\newcommand{\eee}{\end{enumerate}}
\newcommand{\bei}{\begin{itemize}}
\newcommand{\eei}{\end{itemize}}
\newtheorem{theorem}{Theorem}

\newtheorem{lemma}[theorem]{Lemma}

\usepackage{amsfonts}

\def\id{\mathbb{I}}

\def\01{\{0,1\}}

\newcommand{\ket}[1]{|#1\rangle}
\newcommand{\bra}[1]{\langle#1|}

\newcommand{\proj}[1]{|#1\rangle\langle#1|}

\newcommand{\rank}{\operatorname{rank}}

\def\<{\langle}
\def\>{\rangle}
\def\ot{\otimes}
\def\ermax{E_R^{\max}}
\def\esmax{E_S^{\max}}
\def\ecal{{\cal E}}
\def\hcal{{\cal H}}
\def\ecalr{{\ecal_R}}
\def\etaE{\eta_{E-E_S}}
\def\rhor{\rho_R}
\def\rhos{\rho_S}
\def\gibbs{\tau}
\def\ergotropymin{F_\epsilon^{\min}}
\def\ergotropymax{F_\epsilon^{\max}}

\def\rhodec{\omega}
\def\s{\,\,\,\,}
\def\dmin{D_{\min}}
\def\dmax{D_{\max}}

\def\wit{\psi_W}
\def\Hw{\hat{W}}
\def\Hin{H}
\def\Hout{H'}

\def\mainsection{Main Section}
\def\fmin{F_{\min}}
\def\fmax{F_{\max}}
\def\tauout{\tau'}
\def\ep{\epsilon}
   \def\urlprefix{}
   \def\url#1{}
   \def\eprint#1{Preprint at http://arxiv.org/abs/#1}

\begin{document}

\title{Fundamental limitations for quantum and nanoscale thermodynamics}
\author{Micha\l{} Horodecki}
\affiliation{
IFTIA, University of Gda\'{n}sk, 80-952 Gda\'{n}sk, Poland }
\author{Jonathan Oppenheim}
\affiliation{
DAMTP, University of Cambridge, CB3 0WA, Cambridge, UK}
\affiliation{University College of London, Department of Physics \& Astronomy, London, WC1E 6BT and London Interdisciplinary Network for Quantum Science}


%
%
%
%

\begin{abstract}
The relationship between thermodynamics and statistical physics is valid in the thermodynamic limit - when the number of particles becomes very large. Here, we study thermodynamics in the opposite regime - at both the nano scale, and when quantum effects become important. Applying results from quantum information theory we construct a theory of thermodynamics in these limits. We derive general criteria for thermodynamical state transformations, and as special cases, find two free energies: one that quantifies the deterministically extractable work from a small system in contact with a heat bath, and the other that quantifies the reverse process. We find that there are fundamental limitations on work extraction from nonequilibrium states, owing to finite size effects and quantum coherences. This implies that thermodynamical transitions are generically irreversible at this scale. As one application of these methods, we analyse the efficiency of small heat engines and find that they are irreversible during the adiabatic stages of the cycle.
\end{abstract}
\maketitle

\vspace{3mm}
Corresponding author: Jonathan Oppenheim, e-mail address {\tt j.oppenheim$@$ucl.ac.uk} \\
\vspace{3mm}

One of the most basic quantities in thermodynamics is the Helmholtz free energy 
\begin{align}
F(\rho) = \<E(\rho)\> - TS(\rho)
\label{eq:helmfreeenergy}
\end{align}
with $T$ the temperature of the ambient heat bath that surrounds the system, $S(\rho)$ the entropy of the system, and $\<E\>$ 
its average energy.  It tells us whether a system at constant volume 
and in contact with a heat bath can make a spontaneous 
thermodynamical transition from one state to another.  A transition can only happen if the free energy of the final state is lower than that of 
the initial state. 
The difference in free energy between the initial and final state is also the amount of work which can be extracted from a system
in a thermal bath.  It also gives the amount of work required to perform the reverse process, since thermodynamics at the macroscopic scale
is reversible.

However, the free energy is only valid 
in the thermodynamical 
limit -- when $\rho$ is composed of many particles and is classical,
in the sense that it is in a state which is a probabilistic mixture of different energies. 
But thermodynamical effects are not only important in the macroscopic regime -- they are becoming
increasingly important as we probe and manipulate small systems from the micro up to the mesoscopic scale.  
Already, molecular motors and micro-machines
\cite{Scovil1959masers,Geusic1967quatum,Alicki79,howard1997molecular,geva1992classical,Hanggi2009brownian,allahverdyan2000extraction,feldmann2006lubrication,linden2010small}
have been constructed in the lab\cite{Scovil1959masers,scully2002afterburner,rousselet1994directional,Faucheux1995ratchet}
and thermodynamical effects are increasingly important in quantum devices and in the construction of 
quantum computers and memory\cite{Landauer,bennett82,baugh2005experimental}.  Likewise, quantum effects have implications for
thermodynamics~\cite{gemmer2009quantum,popescu2006entanglement,del2011thermodynamic}.

In this article, we derive necessary and sufficient conditions for thermodynamical state-to-state transitions  which are valid even when the thermodynamical limit is not taken, and even when the system is quantum. 
We call these
conditions thermo-majorization. As a special case of this more general result, we derive
two free energies valid in this regime. 
We also quantify the extent to which general state transformations are irrversible, and 
derive a criteria for when transitions between two states block-diagonal in energy eigenbasis can be made
reversible in the micro-regime. We find that there are particular processes which approach the ideal efficiency, provided that certain
special conditions are met. 
Our most basic result concerns the state of a microsystem, which is out of equilibrium, 
and we ask first, how to define microscopic work, and then we provide the optimal 
amount of work that can be drawn from the system when in contact with a heat bath, as well as the amount
of work required for the reverse process (the {\it work of formation}).  
The obtained amount of work is given by a version of the relative entropy 
distance of the state from the Gibbs state.  Similarly, 
the work needed to create a system is given by another version of the relative entropy distance 
to the Gibbs state. These two cases are examples of our full thermo-majorization result which 
includes characterization of all possible transitions between states block diagonal in the energy eigenbasis 
in the presence of a heat bath.

\section{Results}
\subsection{Conceptual prerequisites}
In the macroscopic regime, the standard free energy can be expressed by means of
the relative entropy \cite{donald1987free}, and this can be used to compute the work drawn 
from non-equilibrium states \cite{espositoB-2011,thermoiid}. However, it is surprising that
in the micro-regime, where fluctuations may dominate, the distillable work and work of formation
can also be expressed as relative entropies, albeit very different ones. This is because in the microregime,
one has a single system with large fluctuations, and it is not at all clear that one can draw work deterministically, as
one does in the macroscopic case. 
One might have imagined that one need to look at the non-deterministic case where one sometimes succeeds in drawing work, and
sometimes doesn't. This approach, while certainly of interest, has the disadvantage that without deterministic work
extraction, it can be difficult to separate work from the entropy stored with the work, since if one is not almost certain to draw work,
the work will be inherently noisy. To make the distinction between work and noise, one then invariably
looks at running a thermodynamic cycle many times, and this doesn't allow one to fully consider individual systems.
In contrast, here, we are able to make strong statements about what will happen to a single system.

Our results were possible due to combining a number of existing concepts. The case of manipulation
of entropy, and deterministic transitions when the Hamiltonian is trivial, was undertaken in \cite{uniqueinfo},
where transition criteria and work extraction were given by  majorization conditions. This can be considered a resource theory of ``purity'' or 
entropy. In \cite{dahlsten2011inadequacy},
a probability of failure was allowed for extracting work, allowing work to be quantified by smooth entropies, and this was extended in \cite{del2011thermodynamic}
to the case where one only wants to extract work from one subsystem of a bipartite state, while preserving the other subsystem.
In these three cases, since energy was essentially decoupled from entropy, work extraction was a purely information-theoretic task -- defined as going from a mixed state
to a pure state -- enabling a generalisation
of  Landauer's principle, saying that a pure state of a two level system without a Hamiltonian is equivalent to 
$kT \ln 2 $ of work. Indeed, one links the concept of work to entropy change, simply through Landauer's principle.

However, thermodynamics is not merely the study of entropy, but rather the interplay between energy and entropy. 
Entropy is only half the picture. 
A key tool we will need to use, is a resource theory which combines the resource theory of purity, with that of ``asymmetry'' which
is the study of manipulations constrained by superselection rules\cite{gour_resource_2008}, of which energy conservation is a special case.
Combining these two resource theories allows one to study thermodynamics in all its generality.
A paradigm was presented in \cite{Beth-thermo} and 
we shall employ its components as a resource theory here, but with two new twists - 
first, while keeping the system microscopic, we consider its interaction with a large heat-bath.  This allows us to combine 
the above-mentioned approaches together with the quantum information theory of resources, 
to obtain a novel theory of thermodynamics in the micro-regime.  Secondly, we will add a work system into the picture,
which will allow us to define work as the process of raising an energy level of this work system. The skeleton of our construction 
is the theory of resources, 
and in a parallel paper \cite{thermoiid}, we show how it reconstructs thermodynamics in the macroregime
in the particular case of many identical copies of a micro-system. The present paper concerns itself with the microregime.  
Remarkably, even when
we have a non-trivial Hamiltonian acting on our system, and manipulate systems through a non-trivial interaction Hamiltonian between
the system and reservoir, we still find that work extraction
and formation are given by elegant information theoretic quantities.

\subsection{Thermal Operations}
\begin{figure*}
  \centering
  \begingroup%
  \makeatletter%
  \providecommand\color[2][]{%
    \errmessage{(Inkscape) Color is used for the text in Inkscape, but the package 'color.sty' is not loaded}%
    \renewcommand\color[2][]{}%
  }%
  \providecommand\transparent[1]{%
    \errmessage{(Inkscape) Transparency is used (non-zero) for the text in Inkscape, but the package 'transparent.sty' is not loaded}%
    \renewcommand\transparent[1]{}%
  }%
  \providecommand\rotatebox[2]{#2}%
  \ifx\svgwidth\undefined%
    \setlength{\unitlength}{15cm}%
    \ifx\svgscale\undefined%
      \relax%
    \else%
      \setlength{\unitlength}{\unitlength * \real{\svgscale}}%
    \fi%
  \else%
    \setlength{\unitlength}{\svgwidth}%
  \fi%
  \global\let\svgwidth\undefined%
  \global\let\svgscale\undefined%
  \makeatother%
  \begin{picture}(1,0.22403389)%
    \put(0,0){\includegraphics[width=\unitlength]{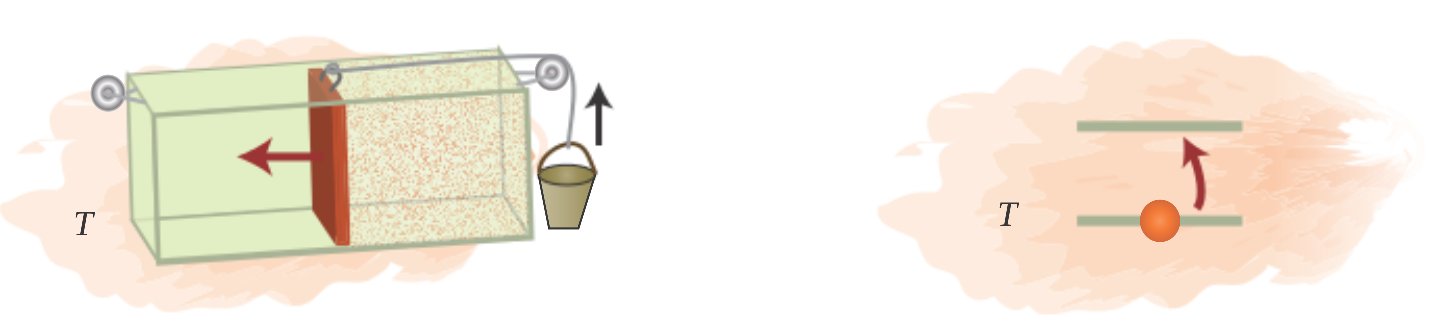}}%
    \put(0.0080507,0.19925785){\color[rgb]{0,0,0}\makebox(0,0)[lb]{\smash{a)}}}%
    \put(0.62142467,0.2002424){\color[rgb]{0,0,0}\makebox(0,0)[lb]{\smash{b)}}}%
  \end{picture}%
\endgroup%
  \caption{Macroscopic and microscopic work. a) A macroscopic heat engine which performs work by lifting a heavy weight a certain height. b) In the quantum
or micro-regime, we can think of work as the ability to excite a two-level system from one energy state to a higher one.  
Having many of these atoms would
allow us to perform macroscopic work -- for example, we could use the atoms in a laser.  An amount of work $W$ can be used to produce
a transition from the state $\proj{0}$, 
to the state $\proj{1}$, with Hamiltonian $\hat{W}=W\proj{1}$ (we call such a two-level system, the {\it work qubit}, or {\it wit}).
We can use such a system as a basic work storage unit, since our results will not depend on what physical system is used.
}
\label{fig:weight}
\end{figure*}
We will first consider a quantum system
\begin{align}
\rho=\sum \sigma(E,E',g,g')\ket{E,g}\bra{E',g'}
\end{align}
with a fixed Hamiltonian $H$ and eigenstates of energy $E$ given by $\ket{E,g}$, in contact with a heat bath.
We are interested in the types of state transitions which are allowed, and in particular, our ability to use the system as a resource to extract work. 
We will then consider the case where the Hamiltonian of the initial and final state is not the same, so that the system undergoes a non-cyclic
evolution.

Instead of considering macroscopic work (the pushing out of a piston, or the raising of a weight), we consider microscopic work -- 
for example, the exciting
of an atom from its ground state to an excited state (Figure \ref{fig:weight}).  We can thus use a two level system to store work.
Because the amount of extractable work can be small, we require precise accounting of all sources of energy.  We thus consider a paradigm where 
extraction of work, and other operations must be done using energy conserving operations\cite{Beth-thermo,thermoiid}, 
so that any energy which is transferred to or from the resource system and heat bath, is transferred from or to the system which stores work.  
We do not impose any additional constraints, since we wish to explore fundamental limitations on what can be accomplished on work extraction and formation.
We call the class of operations that are allowed {\it Thermal Operations} -- a fuller discussion of which is contained 
in \ref{sec:thermo_resource}, including how it is related to other natural paradigms.  
This casts thermodynamics as a {\it resource theory} \cite{Werner1989,BDSW1996,uniqueinfo,thermo-ent2002,Beth-thermo,thermoiid}, which allows us to exploit some mathematical machinery from information theory. Thermodynamics is then viewed as 
a theory involving state transformations in the presence of a thermal bath.
The extraction or expenditure  of work can be included in such a paradigm, because
it is equivalent to a state transformation -- the state of the work qubit is raised or lowered from one energy eigenstate to another.  


Having precisely accounted for all sources of energy, we can apply techniques from {\it single-shot} information theory -- a branch
of information theory specialising in arbitrary resources as opposed to situations where we have many copies of independent and identically distributed
bits of information (see e.g. \cite{Tomamichel-thesis}).  The techniques are thus ideally suited to the case where we want to extract work from a small single system or one whose subsystems are
highly correlated.  It is also applicable when we wish to extract a deterministic amount of work rather than just extract it statistically, 
as we can do here by considering systems in contact with a large heat bath which diminish the effect of statistical fluctuations of the system.

\subsection{Extractable  work}
In this more general
setting, we show in \ref{sec:work}, that  
the quantity which replaces the Helmholtz Free Energy for calculating the extractable
work in the quantum regime is
\begin{align}
\fmin(\rho) = -kT\ln\sum h(\rhodec,g,E_i) e^{-\beta E_i}
\label{eq:ergotropy}
\end{align}
where 
$\rhodec=\sum_{E}P_E\rho_\epsilon P_E$ with $P_E=\proj{E}$
is the state $\rho$ decohered in the energy eigenbasis (i.e. off-diagonal terms are set 
to zero),
$h(\rhodec,g,E_i)$ is $1$ if energy level $\ket{g,E_i}$
is populated and $0$ otherwise. $\beta$ is the inverse temperature, and $k$ is Boltzmann's constant.
For microscopic systems, one can generically extract very little work deterministically without allowing 
a tiny probability $\epsilon$ of failing to draw work\cite{dahlsten2011inadequacy}.
In \ref{sec:work} we consider this situation
and show that a {\it $\beta$-smoothed} version of $\fmin$, called $\ergotropymin$, gives the optimal and 
achievable amount of work extractable from the resource.  It's expression is found in \ref{sec:work}
(Supplementary Equation \eqref{eq:smoothmin})
and in the special case that the Hamiltonian is trivial $H=0$, it corresponds to the expression of \cite{dahlsten2011inadequacy}.

In terms of information theoretic quantities, we can write 
\begin{align}
\fmin(\rho)-\fmin({\gibbs})=kT \dmin(\rho||\gibbs)\s ,
\end{align}
where  $\dmin(\rhodec||\gibbs):=-\ln\tr{\Pi_{\rhodec}\gibbs}$ is 
the min-relative entropy\cite{petz,datta2009min}
with $\Pi_{\rhodec}$ the projector onto the support of 
$\rhodec$ and $\gibbs$ is the Gibbs state $\gibbs=Z^{-1}\sum_{E,g}e^{-\beta E}\proj{E,g}$ with partition function $Z$. 
The min-relative entropy and single-shot free energy has been independently introduced as a lower bound for work extraction from
classical states using a model of a series of independent interactions with a heat bath\cite{aaberg-singleshot}.

In the thermodynamical limit $D_{min}(\rho||\gibbs)$
becomes \cite{thermoiid}  
the relative entropy  $S(\rho||\gibbs):=-\tr\rho\log\gibbs+\tr\rho\log\rho$
which is equal to $F(\rho)-F(\gibbs)$\cite{donald1987free}.
Thus, while the maximum amount of work $W$ which can be extracted when a macroscopic
system is in contact with a heat bath, is 
$W(\rho)=F(\rho)-F(\tau)$,
more generally it is $W=\ergotropymin(\rho)-\ergotropymin(\tau)$
and only in the thermodynamical
limit do we recover the traditional result. 

Although $\fmin$ looks very different to the Helmholtz Free Energy, it can be compared to it easily
in the situation where the given state $\rho$ has energy fluctuations $\delta E$ which are small compared with the average energy $\< E\>$ as is the case with macroscopic thermodynamical systems. 
We then consider a version $\rho_\epsilon$ of the state $\rho$, with the tails of weight $\epsilon$ removed 
(this is more or less what happens when we smooth $\fmin$ as discussed in \ref{sec:work}) 
and find by Taylor expanding $\fmin(\rho_\ep)$ around the mean energy and taking the 
zeroeth order approximation that
\begin{align}
\fmin(\rho_\epsilon) \approx
 E - kT \ln\rank(\rhodec_\epsilon)
\label{eq:mesoergotropy}
\end{align}
We can now compare this with the Helmholtz Free Energy. In the case where the system is block-diagonal in the energy eigenbasis i.e.
\begin{align}
\rho=\sum_{E,g,g'}\sigma_{E,g,g'}\ket{E,g}\bra{E,g'}\,\,\,\, ,
\label{eq:classical}
\end{align}
we have that $\rho=\rhodec$.  
Then, for extensive systems and the case of many particles $n$, the quantity 
$\ln\rank(\rhodec_\epsilon)=\ln\rank(\rho_\epsilon)\approx S(\rho)$ 
with $\epsilon$ going to
zero exponentially fast in $n$. (For example, for many non-interacting subsystems such as an ideal case, we may take the system to be composed of many systems in state 
$\rho^{\otimes n}$.  We then obtain the classical results\cite{thermoiid}, and the smoothed min and max entropies approach the 
von-Neumann entropy\cite{Rennerphd};  for extensive, isotropic systems, correlations don't play a role in thermodynamical quantities, and related results hold.) We then have that
Equation (\ref{eq:mesoergotropy}) approaches the Helmholtz Free Energy.

In general however, $\ln\rank(\rho_\epsilon)$ is larger than the entropy $S(\rho)$, especially in the case where we just have a single 
system in the micro-regime,  meaning that $\ergotropymin$ is smaller than the free energy.  
The finite size of the system means that less
work can be extracted.

There is a second reason why a limitation exists on the amount of extractable work.  A quantum system $\rho$ needn't be in 
the form of Equation (\ref{eq:classical}) and in particular can have off-diagonal terms connecting different energy eigenstates.  
However, it is not $\rho$ which enters into Equation (\ref{eq:ergotropy}), but rather the state $\rho$ decohered
in the energy eigenbasis, namely $\rhodec$.  
Thus, to zeroeth order, rather than the rank of $\rho_\epsilon$ replacing the entropy, it is 
the rank of $\rho_\epsilon$  dephased  in the energy eigenbasis
that replaces the entropy.  This quantity is generally larger than the rank of $\rho_\epsilon$ which is why for systems with quantum
coherences of energy, there is a further limitation on how much work can be extracted.  As an example, consider the pure quantum state
\begin{align}
\ket{\psi}=\sum_{E}\sqrt{\frac{e^{-\beta E}}{Z}}\ket{E}\s .
\label{eq:puregibbs}
\end{align}
It has entropy and rank equal to zero.  However, when dephased in the energy eigenbasis to produce $\rhodec$, it becomes the Gibbs state if
the energy levels are non-degenerate,
and has free energy $-kT\ln Z$; no work can be extracted from it, despite it having zero entropy.  However, as we approach the thermodynamic
limit, the coherences matter less and less, and the
free energy in the quantum case approaches the free energy for classical states\cite{thermoiid}, and again, $\fmin$ approaches the 
Helmholtz Free Energy.

\subsection{Work of formation}
The fact that at the quantum or nanoscale one can't extract the work as given by the free energy, implies that there is an inherent irreversibility
in thermodynamic transformations.  This can also be seen as follows -- the maximum amount of work which can be extracted from a system $\rho$ 
in contact with a heat
bath is given by $F_\ep^{\min}(\rho)-F_\ep^{\min}(\gibbs)$.
In the process, the system is transformed from state $\rho$ to the Gibbs state $\gibbs$.
But if we wish to use work to perform the reverse process, namely transform Gibbs states into 
$\rho$ using work,
then we show in \ref{sec:work} that the amount of work which is required is $\ergotropymax(\rho)-\ergotropymax(\gibbs)$ with
\begin{align}
\ergotropymax(\rho)=
kT\inf_{\rho_{\epsilon}}
\log\min \{\lambda:\rho\leq\lambda\gibbs \}-kT\ln Z
\label{eq:ergotropymax}
\end{align}
in the case where $\rho$ is diagonal in the energy eigenbasis.   Here the infimum is taken over states $||\rho_{\ep}-\rho||\leq \ep$
with the optimal {\it smoothing} given in \ref{sec:work}. In the case where the Hamiltonian is trivial $H=0$, Equation (\ref{eq:ergotropymax})
can be  interpreted as an upper bound on the amount of work which can be extracted~\cite{dahlsten2011inadequacy}, which coincides with the fact that in such a case, we interpret it as the amount of work which was put into creating the state to begin with. Such an interpretation can also 
be given to Equation (\ref{eq:ergotropymax}) in the case of full thermodynamics with energy.
 
Again, to compare this quantity to the Helmholtz Free Energy, it's worth looking at the zeroeth order approximation after expanding in powers
of $\delta E/\langle E \rangle$.  We find 
\begin{align}
\ergotropymax(\rho)\approx \inf_{\rho_{\epsilon}}[\langle E \rangle-kT\ln 1/p_{max}]
\end{align}
where $p_{max}$ is the largest probability.
To zeroeth order, we see that $\ergotropymin$ is related to the $\ln\rank$ of the density matrix, the Helmholtz Free Energy to the entropy
of the density matrix, and $\ergotropymax$ to $\ln(1/p_{\max})$.  When all probabilities of a density matrix are roughly equal, as is the case
for many non-interacting particles $\rho^{\otimes n}$, then these
three quantities are equal as well.  However,
in general, $\ergotropymin\leq F\leq\ergotropymax$, so that at the nanoscale
we can generally extract less work from a resource than is required to create the resource, leading to a fundamental irreversibility in
thermodynamical processes. In terms of information theoretic quantities, $F_{\max}(\rho)-F_{\max}({\gibbs})=T D_{max}(\rho||\gibbs)$,
where  $D_{max}(\rho||\gibbs):=\log\min \{\lambda:\rho\leq\lambda\gibbs \}$ is the max-relative entropy\cite{datta2009min}.  
As we approach the thermodynamic limit  $\ergotropymin\approx \ergotropymax$, and reversibility is restored\cite{thermoiid} 

\subsection{More general thermodynamical transformations}
\begin{figure*}
  \centering
  \includegraphics[width=15cm]{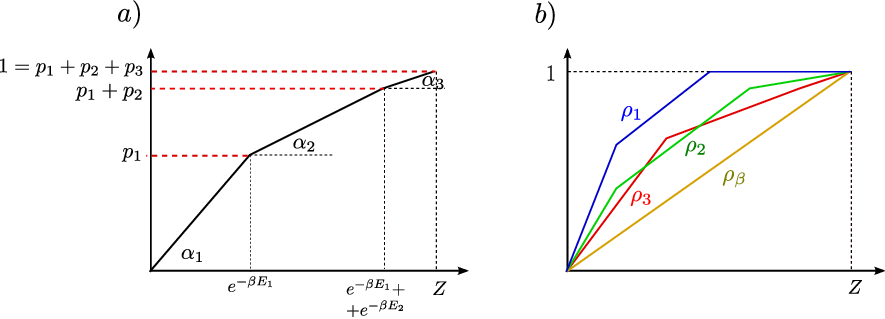}
  \caption{Thermomajorization. Consider probabilities $p(E,g)$ of the initial system $\rho$ to be in the $g$'th state of energy $E$.  Now let us
put $ p(E,g)\e^{\beta E}$ in decreasing order $p(E_1,g_1)\e^{\beta E_1}  \geq  p(E_2,g_2)\e^{\beta E_2}\geq p(E_3,g_3)\e^{\beta E_3} ...$ -- we say
that the eigenvalues are $\beta$-ordered.  We can do the same for system $\sigma$ 
i.e. $\e^{\beta E_1} q(E_1,g_1) \geq \e^{\beta E_2} q(E_2,g_2)\geq\e^{\beta E_3} q(E_3,g_3)...$.  Then the condition which determines whether
 we can transform $\rho$ into $\sigma$ is depicted in the above figure.  Namely, (a) for any state, we construct a curve with points $k$ given
 by $\{\sum e^{-\beta E_i}/Z,\sum_i^k p_i \}$. Then (b) a thermodynamical transition from $\rho$ to $\sigma$ is possible if and only if, the curve of $\rho$ 
lies above the curve of $\sigma$.  One can make a previously impossible transition possible by adding work in the form of the pure state $\psi_W$
which will scale each point by an amount $e^{-\beta W}$ horizontally.}
\label{fig:thermomaj}
\end{figure*}

More generally, we would like to have criteria which tells us whether one
state can be transformed into another under some thermodynamical process.  As we have seen, because of finite size or quantum effects,
the decreasing of the free energy is not a valid criteria which determines whether a thermodynamic transition can occur.   
For transitions between a system $\rho$ and a system $\sigma$, 
both diagonal in the energy eigenbasis, we can derive necessary and sufficient criteria, which we
call thermo-majorization.  It is based on the majorization condition for state transformations which is a necessary and sufficient condition
for state transformations under permutation maps.  Its construction is given in \ref{sec:major}, and we state the result
in Figure \ref{fig:thermomaj}. An alternative derivation of our thermo-majorization condition can be obtained by 
adapting results of Ruch and Mead, studied in the context of decoherence and a 
particular master equation\cite{ruch1975diagram,ruch1976principle,ruch1978mixing} 
and combining them with our proof that Thermal Operations are Gibbs preserving ones given in \ref{sec:gibbspreserving}
(this latter result in the special case
of a heat bath composed of many independent systems was provided in \cite{janzing_thermodynamic_2000}).  
The derivation we present in \ref{sec:major} is more direct, and proves the conjecture that the ``mixing distance'' decreases
in thermodynamical systems -- a problem which has been open since 1975~\cite{ruch1976principle}. We are also able to prove the converse.  

In the case where $\rho$ is not diagonal in the energy eigenbasis, but the final state $\sigma$ is diagonal, then transformations are possible if and 
only if transformations are possible from $\rhodec$ to $\sigma$.  The reason is simple -- dephasing in the energy eigenbasis commutes 
with Thermal Operations\cite{thermoiid} since the latter must conserve energy. 
 Since we can dephase the final state without changing it (as it is already diagonal in
the energy basis) we can use the fact that dephasing commutes with our operations to instead dephase the initial state without changing
whether the transformation is possible.  
  
In the case where the final state is also non-diagonal in the energy basis,
the criteria for which transformations are possible 
depends on the coupling one has with the system, and especially, the degree of control one has of the system.  
Thus far,
our results have not depended on having fine-grained control of the system and heat bath -- the interaction depends on macroscopic variables such
as total energy $E$, but the mapping between microstates $g$ does not matter\cite{thermoiid}.  This is not necessarily the case
during the formation process of states with off-diagonal terms.  Thus, while Equation (\ref{eq:ergotropy}) for the extractable work 
holds in general, 
the same is not true of 
Equation (\ref{eq:ergotropymax}) for the formation process.  This is because for the formation process of transforming Gibb's states into a state $\rho$
which is not diagonal in the energy eigenbasis, it is generally not possible to make such a transformation using Thermal Operations 
without additional resources. 
In the case of formation of many copies $n$ of $\rho$, the additional resource can be two level pure states in a superposition of energy levels\cite{thermoiid},
and the size of the system required scales sublinearly in $n$ and hence vanishes as a fraction of $n$. 

\subsection{Changing Hamiltonians}

So far we have considered transitions between the states of a system with fixed Hamiltonian. This might suggest that our approach does not cover 
the microscopic analogue  of thermodynamical processes between
equilibrium states with different initial and final Hamiltonians\cite{Alicki79}, 
such as isothermal expansions of a gas in a container. Yet, fundamentally, 
a time dependent Hamiltonian 
is only an effective picture of a fixed Hamiltonian of a larger system, and we shall 
show below how to describe such transitions in the microscopic regime.


Namely we introduce a qubit on system $C$ which we can act on to switch the Hamiltonian from $\Hin$ to $\Hout$  
(we call this the {\it switching qubit}).  
We can for example take the total Hamiltonian to be
\begin{align}
H_{{\rm tot}}=\proj{0}_C\otimes \Hin + \proj{1}_C\otimes \Hout + W\proj{1}
\label{eq:changingham}
\end{align}
and take the initial state of the work qubit, switching qubit and system to be $\proj{00}_{CW}\otimes\rho$ and final state to be  
$\proj{11}_{CW}\otimes\sigma$, 
so that we are effectively changing the Hamiltonian acting on $\rho$, and gaining or losing work in the work qubit when we make the transition
to $\sigma$.
We now consider a transition between $\rho$ and $ \tauout$, the thermal state with Hamiltonian $H'$,  and want to know what 
value (positive or negative) for $W$ allows us to make this transition. 

\begin{figure*}
  \centering
  \includegraphics[width=15cm]{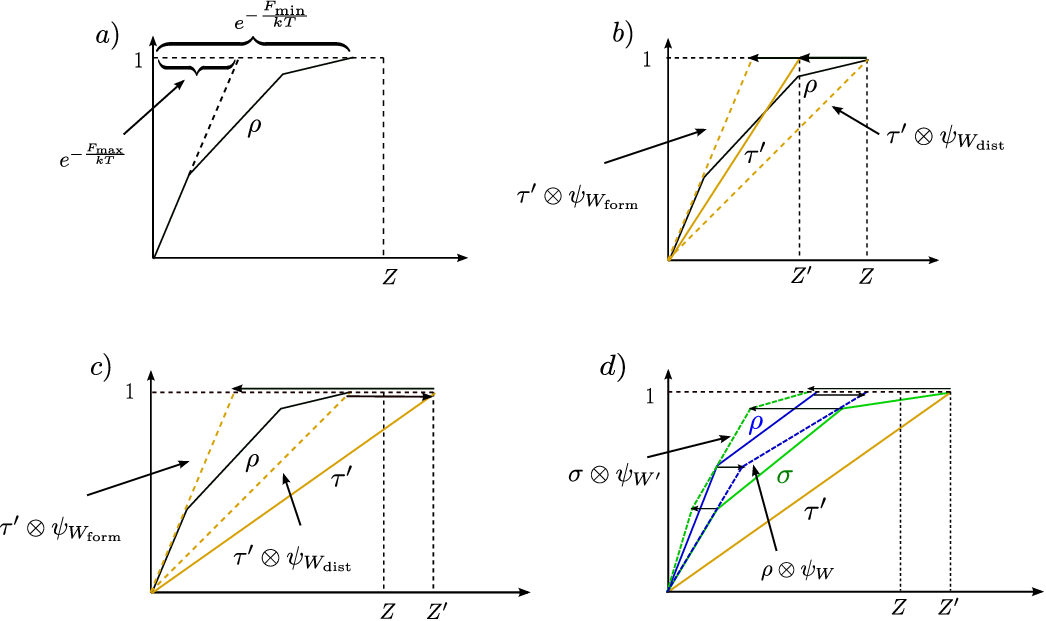}
  \caption{
{\bf Distillable work, work of formation, and the general case.} (a) Graphical representation of two free energies.
  For Gibbs state they coincide. (b),(c): The scenario of changing Hamiltonian we can mimic by adding to the system $S$ ancilla $C$  switching between initial $\Hin$ and final Hamiltonian $\Hout$, with partition functions $Z,Z'$, respectively. We can consider transition $\rho \to \tauout$ and $\tauout \to \rho$,
and obtainable works denote by  $W_{\rm dist}$ and  $W_{\rm form}$ respectively. The work
can be of either signs. Adding/Subtracting  work to a state is graphically represented as changing its slope.  
Formation is depicted by an arrow going from $\rho$ to $\tauout$, distillation by arrow
going from $\tauout$ to $\rho$. The directions of the arrow to the right/left means that the work is positive/negative in a given transition. 
Positive work, means that we obtain work during the process.
We depict two out of possible four cases of work signs:  (b) both works are negative
(c) work of formation is negative while work of distillation is positive. (d) The interconversion of two arbitrary states is depicted by means of adding/subtracting work; $W$ is the maximal work that can be obtained by  
the transition $\rho\to \sigma$; $W'$ is the minimal work needed to perform the transition $\rho\to\sigma$.
Here $\psi_E=|1\>\<1|$ is the excited state of the Hamiltonian $H=E|1\>\<1|$, for $E=W,W',W_{\rm form}, W_{\rm dist}$.}
%
%
\label{fig:changingham}
\end{figure*}

The results, obtained by means of thermo-majorization are depicted in Figure \ref{fig:changingham}. One finds
\begin{align}
W=\ergotropymin(\rhodec)-\ergotropymin(\tauout)
\label{eq:gmin}
\end{align}
for extracting work, and  for the amount of work required to form $\rho$ (provided it is diagonal 
in energy eigenbasis) from the thermal state, 
we obtain
\begin{align}
W=\ergotropymax(\rho)-\ergotropymax(\tauout)
\label{eq:gmax}
\end{align}
This result does not depend on the form of the Hamiltonian of Equation \eqref{eq:changingham} -- we only require that
at late times, there is no interaction between the work qubit and the other systems (since we need to be able to separate out 
the work qubit to use in some future process). More general state-to-state transformations assisted by work are also depicted.

To derive Equations \eqref{eq:gmin}-\eqref{eq:gmax}, 
we $\beta$-order the $p_i$ and $q_i$ corresponding to $\rho\otimes\proj{00}$, and $\sigma\otimes\proj{11}$ respectively.  Then  
the thermo-majorization coordinates $k$ of $\rho\otimes\proj{00}$ are given by $\{ \sum_1^k e^{-\beta E_i}, \sum_1^k p_i \}$, 
and those of $\sigma\otimes\proj{11}$ are $\{ \sum_1^k e^{-\beta (E'_i+W)}, \sum_1^k q_i \}$.
The thermo-majorization condition for a transition is that for all $k$, the points associated with $\rho$ are above that of $\sigma$
and they take a particularly simple form when either $\rho$ or $\sigma$ is the thermal state.  These two cases are shown 
in Figure \ref{fig:changingham}.  The case where the final state is thermal for Hamiltonian $\Hout$,  $\sigma=\tauout$, 
and the work qubit is excited corresponds to distillation, since 
no further work can be drawn for fixed $\Hout$ once the state is thermal, and a transition to another state can always be followed by
a transition to the thermal state. Therefore drawing work by relaxing the state to a thermal state is completely general, and gives us
Equation (\ref{eq:gmin}).  If $\rho$ has off-diagonal terms, then the distillable work is given by the decohered version $\rhodec$ in Equation
(\ref{eq:gmin}), due to the same reasoning as we used earlier -- the final state is simply the work qubit, since everything else can be thrown 
away, and therefore is diagonal in the energy eigenbasis. Since decohering the final state doesn't change the final state, and  decohering 
with respect to the total Hamiltonian commutes with Thermal Operations, we can do it to the initial state without affecting the amount of work
extractable.

The case where we adjust $W$ so that $\rho\otimes\proj{00}_{CW}$ is thermo-majorized by $\sigma\otimes\proj{11}_{CW}$ gives us the 
formation process, and free energy of Equation (\ref{eq:gmax}).   
The case where both initial and final states $\rho$,$\sigma$
are thermal is also depicted in Figure \ref{fig:changingham}, and leads to the ideal classical result, namely 
that a transition is possible if and only if
\begin{align}
W = - kT \ln{Z/Z'}
\label{eq:thermaltransitions}
\end{align}
i.e. the work is given by the difference of standard free energies \eqref{eq:helmfreeenergy}. 

\section{Discussion}
Equation \eqref{eq:thermaltransitions} is a very different result to Equation \eqref{eq:ergotropy}, where we had no ancillary system isolated from the heat bath as in 
Equation (\ref{eq:changingham}).
It shows that for thermal equilibrium states there can be reversibility in some thermodynamical processes, provided they are between
two thermal equilibrium states and the Hamiltonian changes.  In the picture 
of a fixed Hamiltonian, this required at least 
one additional system (the switching qubit), which is effectively not in contact with the heat bath, and we do not draw the maximal amount 
of extractable work from the  total working body, given by 
$\ergotropymin(\rho_S\ot \proj{0}_C)$.  The final state is thermal only on a subsystem $S$ and therefore the amount of drawn work is not maximal. 

This strongly suggests that if we wish to carry out a Carnot cycle to extract work between two heat baths at different temperatures, 
then to get optimal efficiency during the isothermal process, we will need a working body of at
least dimension $2\times 3$.  The first two level system acts as the working body which interacts with the heat baths, while the additional three level
system is needed if we want to switch between different Hamiltonians in order to achieve the optimal isothermal work extraction given by 
Equation \eqref{eq:thermaltransitions}.  Even then, we find that while the two isothermal processes can be made ideal, the two adiabatic processes 
result in additional entropy production, meaning that the Carnot efficiency is not reached over a small number of cycles.  This is analysed in 
\ref{sec:carnot}.

In general, we only get reversibility if there exists a $W$, such that the 
thermo-majorization plot of the initial state $\{\sum e^{-\beta E_i}/Z,\sum_i^k p_i \}$, can get mapped onto the plot of the final
state $\{\sum e^{-\beta (E'_i+W}/Z',\sum_i^k q_i \}$.  Thus reversibility requires a very special condition.  
It is this lack of reversibility
which requires two free energies. There is a connection here with other resource theories. Consider the set of states which are preserved
under the class of operations -- in entanglement theory, these are separable states, and for Thermal Operations, we show in \ref{sec:gibbspreserving} that it is the
Gibbs state. Now, if the theory is reversible, then under certain conditions, the relative entropy distance to the preserved set is the unique
measure which governs state transformations\cite{thermo-ent2002,BrandaoPlenio2007-separable-maps}. For Thermal Operations, the relative entropy distance to the Gibbs state is precisely the  free energy difference\cite{donald1987free}.  Here, in the case of finite sized systems, we see that although we don't have reversibility, the relative entropy distance  to the
preserved set again
enters the picture, but it is the min and max relative entropy. These quantities are monotonically decreasing under the class of Thermal Operations, and provide two measures for state transitions.

\section{Methods}
The proofs are contained in the Supplementary Information. In \ref{sec:thermo_resource}, we case thermodynamics as a resource 
theory, and in \ref{sec:major},
show that the condition for state transformations is given in terms of majorization. In \ref{sec:pure} we consider transitions to 
and from pure states
which we then use in \ref{sec:work} to derive the extractable work, and work of formation. \ref{sec:trumping} discusses the case 
where we allow the use of ancillas, and \ref{sec:gibbspreserving} characterises Thermal Operations, and looks at possible 
transitions in two and three level systems in the case where there are coherences between energy levels. \ref{sec:carnot} 
discusses the details of a small engine undergoing a Carnot cycle.

\section{Author Contributions}
Both authors contributed equally to this work.
\\

{\bf Acknowledgements} 
We thank Robert Alicki, Charles Bennett, Fernando Brandao, Sandu Popescu and Joe Renes 
for discussions. We thank Lidia del Rio for Figure 1, and comments on our draft.
JO is supported by the Royal Society. MH is supported by Polish Ministry of Science and Higher Education grant N N202 231937 and by EC IP QESSENCE. M.H. also acknowledges the support by Foundation for Polish Science TEAM project 
cofinanced by the EU European Regional Development Fund for preparing the final version of this paper.
MH acknowledges the hospitality the of Quantum Computation group at DAMTP, and JO
thanks the National Quantum Information Centre of Gdansk and the Dale Farm Residents Association 
for their hospitality while the manuscript was being completed.





\renewcommand\thesection{Supplementary Note \arabic{section}}
\renewcommand\theequation{S\arabic{equation}}

\renewcommand\figurename{}
\renewcommand\thefigure{Supplementary Figure S\arabic{figure}}
\renewcommand\tablename{}
\renewcommand\thetable{Supplementary Table S\arabic{table}}

\setcounter{figure}{0}
\setcounter{equation}{0}
\setcounter{section}{0}
\setcounter{table}{0}





\section{Thermodynamics as a resource theory}
\label{sec:thermo_resource}

\pagenumbering{gobble}
In the micro-regime, when the amount of work which can be extracted might be of the order of $kT$, we need to
very precisely define what we mean by work, and what processes are allowed during the extraction of work from
a system. For our purposes, obtaining work $\Delta W$ means to obtain an eigenstate of the Hamiltonian  with energy $W_{\rm out}$
starting from an eigenstate of energy $W_{\rm in}$, where $W_{\rm in}-W_{\rm out}=\Delta W$. 
In our approach it will turn out, that the amount of work we can extract from a given system 
does not depend on the Hamiltonian of the system which stores the work, and the particular levels we choose.
We can thus consider a system of the smallest dimension, which carries work $W$. 
This is a two level system with Hamiltonian $\Hw=W |1\>\<1|$. We shall call this a {\it work qubit} (in short, a {\it wit)},
and let $\ket{\wit}$ denote the excited state $\ket{1}$ with energy $W$. 
This is the most economical way of storing work.

Since drawing or adding work can be represented as a state transformation, it is natural to consider 
thermodynamics as a resource theory. Namely, one considers some class of operations,
and then asks how much of some resource can be obtained.  Recent examples of such theories include entanglement theory [39,28],
thermodynamics with no Hamiltonian [22], thermodynamics of erasure [25]
and operations which respect a symmetry [24,40].  
Here, we use the class of operations which
corresponds to thermodynamics [25,21], and then ask by how much we can excite a system 
initially in a pure ground state.
It can be shown that there are a number of equivalent ways
of describing this class of operations [21].  

Since we are interested in extracting work in the presence of a heat bath, one starts by allowing
a free resource of a heat bath at temperature $T$, and with Hilbert space $\hcal_R$. The heat bath is in a 
Gibbs state $\gibbs$, with arbitrary Hamiltonian
and we further allow the addition of any auxiliary system $S'$ with Hamiltonian $H_{S'}$ 
in a Gibbs state. 
Without loss of generality, we can take the initial Hamiltonian to be non-interacting at very early times between the reservoir $R$ and  
the system of interest $S$, as well as any ancillas.  We also want that initially (and finally), the work qubit
is not interacting with the rest of the system, since we want to be able to store the work, and use it in some other process.
We thus have initially $H_{\rm tot}=H_R+ H_S +H_{S'} + \Hw$. 

We now require that all manipulations conserve energy.  This ensures that all sources of work are properly accounted for, and that
external systems are not adding or taking away work.
The dynamics can be implemented by an interaction Hamiltonian, however, if we wish to maintain a precise accounting of all energy,
then the interaction term needs to vanish at the beginning and end of the protocol, otherwise it allows us to pump work into the system at no cost.
Essentially we need to ensure conservation of total energy. This also means that if we wish to model 
a time-dependent Hamiltonian, we should do so by means of a time-independent Hamiltonian with a clock included in the system.  It is not
difficult to show [21], that all of these paradigms which conserve energy, are equivalent to  unitary  transformation 
commuting with the total Hamiltonian.  Essentially, since accounting for all sources of energy requires that the initial and final 
energies are the same, the dynamics must map eigenstates of the Hamiltonian to eigenstates with the same energy.  This is equivalent to considering a fixed Hamiltonian, and  allowing operations which commute with the Hamiltonian.
We also allow discarding subsystems (partial trace). We call this class - {\it Thermal Operations}.   

Note that this paradigm allows one to include different initial and final Hamiltonians as in the example discussed in the \mainsection
\begin{align}
H_{\rm tot}=\proj{0}_C\otimes \Hin + \proj{1}_C\otimes \Hout + W\proj{1}
\label{eq:changingham2}
\end{align}
Via a similar mechanism, one can include interacting terms which vanish at early and late times.
Similarly, the application of a unitary during some time period can be made via application of a fixed Hamiltonian [21]
and using an ancilla which acts as a clock.
Generally, we are interested in transitions between $(\rho_S, H_S)$ and $(\sigma_{S'}$ and $H_{S'})$
(extracting work will be a special case of such a transition). Since in the described approach, 
the Hamiltonian is fixed, such a transition means actually  
$(\rho_S\ot\tau_{S'}, H_S+H_{S'})\to (\tau_{S} \ot \sigma_{S'}, H_S+H_{S'})$
where we have the same initial and final Hamiltonian. 
%

We will here, generally write an arbitrary transition as just  
$(\rho_S, H_S)\to (\sigma_{S}, H_S)$ since the system $S$ can include a number of components
including a clock, a working body, and various other ancillas and coupling systems. Likewise, $H_S$ could include
various coupling terms. We then derive necessary and sufficient criteria for a transition to be possible.
However, one might want to derive conditions for a transitions on some system, while optimising over all 
configurations of the ancillas and working body.
We do that in more detail in [41], where we show to what extent and under what conditions our 
formulas for work extraction and work of formation are robust under such an optimisation. Here we recall 
the main conclusions. The most general scenario is the following: apart from the systems considered so far 
(i.e. heat bath, working body, and work collector) we have also clock and allow any ancillas.
This includes the system that allows us to switch on and off any interaction Hamiltonian between the
various systems. 
The total systems evolve altogether under some time-independent Hamiltonian.
Now, to comply with Planck's formulation of second law, we should at the end of a cycle, not 
change the environment. This means that the state of the clock and other ancillas should be returned 
intact. This latter condition is however too stringent, since in such case the clock would not 
be able to perform its task of switching on and off the interaction. Therefore, we have to allow 
for returning ancillas close to their original state in some distance measure. This then, is
the scenario of catalytic transformations, something which has been studied in the case of
the resource theory of entanglement [42]. 
We discuss this in [41] and give the results in \ref{sec:trumping},
showing that it does not affect our formulas for the min and max free energy for some natural choice of distance measure.


In what follows, we do not consider additional restrictions on the class of operations -- 
we allow any Hamiltonian, and any couplings.  Thus our work provides fundamental limitations to thermal transitions in the lab. 
One might want to consider a more limited class of operations, or add in various practical considerations depending on the physical situation, 
such as restricting the degree of control one has over the resources, or adding in additional couplings between our heat engines and the thermal baths. However, we have found  [21] (see also unpublished results, M. Horodecki and J. Oppenheim)
that the operations needed to distill work can be crude, without fine-grained control.  Nonetheless, for some Hamiltonians, one might demand even less control, in which case, we provide fundamental limitations, rather than matching limitations and achievability bounds.  The same is true for the case where one imposes a further restriction on the class of operations of only coupling the system to a very small reservoir, or small part of the reservoir. In such a case, our fundamental limitations are still of course respected, but the bounds might
not be achievable, in part because sampling a region of a reservoir which is small compared with the length of interactions results in the
system seeing an effective temperature or non-thermal state, rather than the true temperature [43-45].

\subsection*{Assumptions on heat bath, and its relation to the system}
We can assume that Hamiltonians of all systems of concern 
(i.e. heat bath Hamiltonian, auxiliary systems, the resource system itself) have minimal energy zero. 
Let $E_R$ be the energies of reservoir, and $E_S$ be the energies of the system.  
Let $\ermax$, and $\esmax$ be the largest energy of the heat bath and system, respectively
(of course a typical heat bath will have $\ermax=\infty$). 

Our heat bath will be large, while our resource states will be small.
This means that the system Hilbert space will be fixed, while the 
energy of the heat bath (and other relevant quantities such as size of degeneracies) will tend to infinity. 

We now make some assumptions concerning the state and Hamiltonian of the heat bath and then justify them. 
The heat bath is in a Gibbs state with inverse temperature $\beta$. Moreover there exists 
set of energies $\ecalr$ such that
the state of the heat bath occupies energies from $\ecalr$ with high probability, i.e.  
for the projector $P_\ecalr$ onto the states with energies  $\ecalr$ we have
\be
\tr P_\ecalr \rhor\geq 1-\delta 
\ee
and it  has the following properties:
\bee[(i)]
\item The energies $E$ in $\ecalr$ are peaked around some mean value, i.e. 
they satisfy $E\in \{\<E\>- O(\sqrt{\<E\>}),\ldots \<E\>+O(\sqrt{\<E\>})\}$ 
\label{item:peaked}
\item For $E\in\ecalr$ the degeneracies $g_R(E)$ scale exponentially with $E$, i.e. 
\be
g_R(E) \geq e^{c E} 
\ee
where $c$ is a constant.
\label{item:g_exp}
\item For any three energies  $E_R,E_S$ and $E_s'$  such that $E_R\in\ecalr$ and 
$E_S$, $E_S'$ are arbitrary energies of the system, there exist $E'_R \in\ecalr$  
such that $E_R +E_S = E_R'+E_S'$. 
\label{item:matching}
\item For $E\in\ecalr$ the degeneracies $g_R(E)$ satisfy $g_R(E-E_S) \approx g_R(E) e^{-\beta E_S}$, 
or more precisely:
\be
\left|\frac{g_R(E)e^{-\beta E_S'}}{g_R(E-E_S)}-1\right|\leq \delta
\label{eq:g-property}
\ee
for all energies $E_S$ of the system $S$. 
\label{item:g-property}
\eee
{\it Discussion of assumptions:}
\bei
\item[Ad. \eqref{item:peaked}] This is a standard property of a heat bath.
\item[Ad. \eqref{item:g_exp}]  Follows from the condition \eqref{item:peaked}  of small fluctuations 
combined with extensivity of energy. 
\item[Ad. \eqref{item:matching}] Follows from continuity of the spectrum of the heat bath, which is usually the case.
\item[Ad. \eqref{item:g-property}] Follows from 
\begin{align}
g(E+\Delta E)&=e^{S(E+\Delta E)}\nonumber\\
&\approx e^{S(E)+\Delta E \frac{\partial S(E)}{\partial E}}\nonumber\\
&=g(E) e^{\beta \Delta E}
\end{align}
with $S(E):=\ln g(E)$. and $\beta:=\frac{\partial S(E)}{\partial E}$.
\eei 
It is also easy to see that a product $\tau^{\ot n}$ of many copies of independent 
Gibbs states satisfies the above assumptions. 

\subsection*{Notation and preliminary facts}
\label{sec:prelim}
We shall now need a bit of notation.  Let us define $\ideta^X_E$ as a state of a system X proportional to 
the projection  on to a subspace of energy $E$ (according to the Hamiltonian $H_X$ on this system). 
In particular,  $\etaE$ is given by
\be
\etaE=g(E-E_S)^{-1}\sum_g |E-E_S,g\>_R\<E-E_S,g|
\ee
where $g=1,..,g(E-E_S)$, i.e. $\etaE$ is the maximally mixed state of the reservoir with 
support on the subspace of energy $E_R=E-E_S$. 
We shall also use notation $\eta_K=\id/ K$ where the identity acts on a $K$ dimensional space. 

Let us note that the total space $\hcal_R\ot \hcal_S$ can be decomposed as follows 
\be
\label{eq:Hoplus}
\hcal_R\ot \hcal_S= \bigoplus_E \left(\bigoplus_{E_S} \hcal^R_{E-E_S} \ot \hcal^S_{E_S} \right)
\ee
(here for $E\leq E_S$ and $E\geq \ermax +\esmax$ the summation over $E_S$
is suitably  constrained, however we are interested only in energies $E_R$ from $\ecalr$, 
hence these cases will not occur).

Consider an arbitrary state $\rho_{RS}$ which has 
support within $\esmax\leq E\leq \ermax$. We can rewrite it as follows
\be
\rho_{RS}=\sum_E \sum_\Delta P_E \rho_{RS} P_{E+\Delta}\\
\ee
Here $\Delta= -\esmax, \ldots , \esmax$.  The blocks 
$P_E \rho_{RS} P_{E+\Delta}$ we can further divide into sub-blocks 
\be
P_E \rho_{RS} P_{E+\Delta}=
\sum_{E_S\in I_\Delta} \id_R \ot P_{E_S}  P_E \rho_{RS} P_{E+\Delta} \id_R \ot P_{E_S+\Delta}
\ee
where $I_\Delta=\{0,\ldots, \esmax-\Delta\}$ for $\Delta \geq 0$ and 
$I_\Delta=\{-\Delta,\ldots, \esmax\}$ for $\Delta \leq 0$.
The sub-blocks map the Hilbert space $\hcal^R_{E-E_S}\ot \hcal^S_{E_S+\Delta}$ onto  
$\hcal^R_{E-E_S}\ot \hcal^S_{E_S} $   

We can then extract the state $\rhos$ 
\be
\rhos=\sum_{E_S,E_S'} P_{E_S} \rhos P_{E_S'}
\ee
as follows:
\be
P_{E_S} \rhos P_{E_S'} = \sum_{E} \tr_{\hcal^R_{E-E_S}} 
(P^R_{E-E_S} \ot P_{E_S}  P_E \rho_{RS} P_{E+E_S'-E_S} P^R_{E-E_S} \ot P_{E_S'})
\ee
We then have the following technical result that will be a basis for most of our derivations:
\begin{theorem}
\label{thm:oplus_E}
We consider the set of energies
\be
\ecal=\{E:E-E_S\in \ecalr\}
\ee
where $\ecalr$ satisfies assumptions \eqref{item:peaked}, \eqref{item:g_exp} and \eqref{item:matching} listed above. Then 
\be
\forall{E\in \ecal} \quad 
||\frac{1}{p_E}P_E \rhor\ot\rhos P_{E+\Delta} - \oplus_{E_S}\etaE\ot P_{E_S} \rhos P_{E_S+\Delta}||\leq 2\delta
\label{eq:oplus_E}
\ee
and
\be
\sum_{E\in\ecal} p_E\geq 1 -2\delta
\ee
where $p_E=\tr(P_E \rhor \ot \rhos)$.

\end{theorem}
{\it Proof.} Here we sketch the proof for $\Delta=0$. For $\Delta\not=0$ 
the proof is similar. 
Let us fix an energy block $E$. 
Let $E_R=E-E_S$. The state $\tau_R \ot \rhos$  restricted to 
the energy $E$ block is given by 
\be
P_E \tau_R \ot \rhos P_E=\frac{1}{Z_R}\sum_{E_S} e^-{\beta E-E_S } g_R(E-E_S) \id_R^{E-E_S} \ot P_{E_S} \rhos P_{E_S} 
\ee
where $Z$ is the partition function for system $R$, and  $\id_R^{E-E_S}$ is the identity on the subspace $\hcal_R^{E-E_S}$, see \eqref{eq:Hoplus}.
Using \eqref{item:g-property} we have $g_R(E-E_S)=g_R(E) e^{-\beta E_S}$ we get 
\be
P_E \tau_R \ot \rhos P_E \approx \frac{1}{Z_R} e^{-\beta E} g_R(E) 
\sum_{E_S} \frac{\id_R^{E-E_S}}{g_R(E)e^-{\beta E_S}}\ot P_{E_S} \rhos P_{E_S}
\ee
We then use $\frac{\id_R^{E-E_S}}{g_R(E)e^-{\beta E_S}}=\etaE$ and if we drop the prefactor, 
the state is normalised, hence we obtain the claim.

\section{Transformations of classical states: condition in terms of majorization}
\label{sec:major}
Here we will provide a necessary and sufficient condition for transforming the
diagonal part of a density matrix of one state into the diagonal part of another state acting on the same system.
The condition will be in terms of the so called {\it majorization} condition, and it will be necessary and sufficient for
state transformations of classical states (i.e. diagonal in the energy eigenbasis).
The result is contained in Theorem \ref{thm:major} 

\begin{theorem}
\label{thm:major}
Consider two states $\rhos$ and $\sigma_S$ block-diagonal in the energy eigenbasis, on a system
with Hamiltonian $H_S$. The transition $(\rhos,H_S) \ot (\sigma_S,H_S)$ by means of Thermal Operations 
is possible if and only if the state 
\be
\oplus_{E_S}\etaE\ot P_{E_S} \rhos P_{E_S}
\ee
majorizes
\be
\oplus_{E_S}\etaE\ot \sigma_{E_S}
\label{eq:thm-major-sigma}
\ee
for $E$ large enough. Moreover, if the above majorization relation holds
for two states $\rhos$ and $\sigma_S$ not necessarily diagonal in the energy eigenbasis, 
then there exists $\sigma_S'$ such that for all $E_S$ 
$P_{E_S}\sigma_S P_{E_S}=P_{E_S}\sigma_S' P_{E_S}$, and the transition $(\rho_S,H_S) \to (\sigma_S',H_S)$ 
is possible.
\end{theorem}

The majorization condition  reads as follows: 
we have two sets of eigenvalues put in decreasing order $\{\lambda_i\}$ and $\{\lambda'_i\}$,
and we say that $\{\lambda_i\}$ majorizes $\{\lambda'_i\}$ when 
\be
\sum_{i=1}^l \lambda_i \geq \sum_{i=1}^l\lambda_i'
\label{eq:major}
\ee
for all $l$. We say that $\rho$ majorizes $\sigma$ if the eigenvalues of $\rho$ majorize the
eigenvalues of $\sigma$.  

{\it Proof of Theorem \ref{thm:major}}.
The operations that we can perform on the total system, are arbitrary unitary within 
subspaces of fixed energy. From the expression \eqref{eq:oplus_E} it follows that a subspace of fixed energy $E$  contains only the  part of $\rhos$ that is block-diagonal in energy of the system $S$:
\be
P_E \rhor\ot \rhos P_E \approx \oplus_{E_S}\etaE\ot P_{E_S} \rhos P_{E_S}
\label{eq:eta_form}
\ee
Let us denote by 	$\rho_S^E$ the state resulting from applying unitary $U_E$ 
to the above state, and tracing out the heat bath. We shall argue, that 
if $E$ is large enough, then one can choose such unitary $U_E$ that
the resulting $\rho_S^E$ will be the same, as if we had applied to state \eqref{eq:eta_form}
an arbitrary mixture of unitaries. Before we prove it, let us continue with the proof of the theorem. 
Namely, on one hand, by applying $U_E$ to each subspace,
and tracing out the heat bath, we obtain an output state of the form  $\sum_E p_E \rho_S^E$ i.e. 
it is a result of a mixture of operations  applied to each single subspace. 
However, as we've stated, already in a single subspace, we 
can perform an operation which will have the same effect as an arbitrary mixture of unitaries  
applied to the total state followed by tracing out the heat bath. 

Thus we obtain that the most general transition of the block-diagonal part of $\rho_S$ 
can be described by a mixture of unitaries applied to the state \eqref{eq:eta_form}
and a partial trace over the heat bath. Hence, according to [46], 
which says that we can transform a state into 
another state by a mixture of unitaries if and only if the first one majorizes the second one, we can obtain an arbitrary output state  
that is majorized by the state \eqref{eq:eta_form}.  Now, we note that we can further transform such an output state into 
a state of the form \eqref{eq:thm-major-sigma}  without changing the state of subsystem $S$. This is done by a "twirling" 
operation (which is itself some mixture of unitaries):  For each fixed $E_S$ we apply a random unitary to the heat bath part, 
and identity to the part $S^E$. After twirling, any state becomes of the above form. 
This proves the first part of theorem. 

The second part follows from the fact that 
the blocks of fixed total energy contain only the block-diagonal part of $\rhos$, 
and therefore, if we start with a state which is not diagonal, 
then we do not know what happens to off-diagonal terms but, by the above discussion, we do
know how the diagonal part is transformed. Thus, if state \eqref{eq:eta_form} majorizes 
a state of the form \eqref{eq:thm-major-sigma}, then the diagonals of $\rhos$ will be transformed 
into that of $\sigma_S$. 

Finally, we need to prove the claim that we can effectively implement any mixture of unitaries, 
within a single subspace of fixed energy.  To see this, note that we can tensor out a maximally mixed state of size independent of both $E_S$ and $E$,  
and apply unitaries conditioned on the maximally 
mixed state. We do that by 
writing  
\be
\etaE=\frac{\id_K}{K} \ot \frac{\id_{K'_{E_S}}}{K'_{E_S}} 
\ee
where $K'_{E_S} = g(E-E_S)/K$.
Due to our assumptions about the heat bath, the degeneracy of each energy state is exponentially large in energy, 
so we can take such  $K$ that both $K$ and $K'$ are exponentially large in energy.
Thus a given fixed energy block $E$ can be represented as a tensor product of two systems 
$R_1^E$ and $R_2^E S^E$ in a state
\be
\frac{1}{p_E} P_E \rhor\ot \rhos P_E \approx \frac{\id_K}{K} \oplus_{E_S}\frac{\id_{K'_{E_S}}}{K'_{E_S}}
\ot P_{E_S} \rhos P_{E_S}
\ee
We then know [22] that any mixture of unitary transformations 
can be performed on the system $R_2^ES^E$, provided $K$ is large with respect to $K'_E$,
and we shall choose $K$, and the size of the total system, in such a way,
that this is so, and at the same time $K'_E$ can be large too, which we will need further.

Now, the state with $\frac{\id}{K}$ tensored out does not actually differ much from 
the state  $\eqref{eq:eta_form}$, as we anyway will take the system $R_2^E$ 
to be large. Thus the output state coming from a mixture of unitaries performed on 
the state with and without $\frac{\id}{K}$ being tensored out 
have the same effect on the final form of the state of the system $S$.  
This ends the proof of theorem.

Note that in the proof we have used the  assumptions (i-iii) about the heat bath but not (iv).
The latter will be used when we will need to get rid of the heat bath in the majorization expressions in the next section. 

Finally, it is intuitively obvious, that if we add to a heat bath a small system in a Gibbs state, 
this is again a larger heat bath, i.e. it still satisfies our assumptions. 
Indeed, consider a heat bath $R$ which satisfies assumptions (i-iv), 
and another system $S'$, and consider the total Hamiltonian being a sum of Hamiltonians $H_R$ +$H_{S'}$. 
The tensor product of two Gibbs states is a Gibbs state of the total system. 
Since the original heat bath is large, and our system is small, 
then the conditions (\ref{item:peaked}), (\ref{item:g_exp}) and (\ref{item:matching}) are obviously satisfied. 
Then, writing 
\be
g^{RS'}(E) =\sum_{E_{S'}} g^R(E-E_{S'}) g^{S'}(E_{S'})
\ee
and  using the property (\ref{item:g-property}) of $g^R$ one gets 
\be
g^{RS'}(E)\approx g^R(E) Z_{S'}
\label{eq:g_r_sprim}
\ee
where $Z_{S'}$ is the partition function for $S'$. 
This implies, in particular, that  $g^{RS'}$ also satisfies the condition  (\ref{item:g-property}). 

This  proves the following  intuitively obvious lemma:  
\begin{lemma}
\label{lem:add_tau}
Transition between $\rho_S\ot \tau_{S'}\to \sigma_S\ot \tau_{S'}$
is possible if and only if transition $\rho_S\to \sigma_S$ is possible.
\end{lemma}

Thus adding a system in a Gibbs state makes sense only, if we consider transition 
between systems with different Hamiltonian. Then we bring in a system in a Gibbs state, 
only in order to have that Hamiltonian in future processes e.g. we might then transform the Gibbs state
into another state which needed to have that Hamiltonian.

The conditions given thus far for state transformations are all that is needed to draw the full amount
of work from a state, or to form a state from a heat bath.  
This is done in  \ref{sec:pure}
For the remainder of this section, we continue with more general state transformations.

\subsection*{Thermo-majorization}
\label{sec:thermo_major}
We shall now provide an efficient method of finding, whether a transition  $(\rho,H) \to (\sigma,H)$ 
is possible, for states which commute with Hamiltonian $H$. 
%
%
The condition of transformations of the diagonal part of a density matrix given by Theorem \ref{thm:major} 
in terms of majorization involves not only the state, but also the heat bath, hence it is not always 
directly useful. We shall now express the condition given 
by majorization in terms  of the states of system $S$ themselves which will result in an efficient 
algorithm to decide whether a transition between two diagonal states is possible or not.  Essentially, we need to write the eigenvalues
of the state and heat bath, in terms of eigenvalues of only the state.
We shall assume that our input state and output states are diagonal 
in their energy bases, however, even if they are not, the condition we derive determines possible transformations of the diagonal part
of the density matrix, thus the condition becomes necessary, but ceases to be sufficient.

Let $p_{E_S,g}$  be eigenvalues of $\rho$ and $q_{E_S,g}$ be eigenvalues of $\sigma$. 
Then, due to proposition \ref{thm:major} and the condition \eqref{item:g-property}, the state 
$P_E \rhor\ot \rhos P_E$ after normalisation  is close to the state having  the following eigenvalues:
\be
e^{\beta E_S}\frac{p(E_S,g)}{g_R(E)}
\label{eq:eig}
\ee
with multiplicity $g_R(E) e^{-\beta E_S}$, where $E_S$ runs over  all 
energies of the system, and $g$ runs over degeneracies.
Similarly, $P_E \rhor\ot \sigma_S P_E$ has eigenvalues $e^{\beta E_S}\frac{q(E_S,g)}{g_R(E)}$
with the same multiplicity.

The eigenvalues are very small, and they are collected in groups, where they are the same, 
hence the majorization amounts to comparing integrals. 
If one puts eigenvalues into decreasing order, one obtains a stair-case like function,
and majorization in this limit will be to compare the integrated functions (which are then 
piece-wise linear functions).

To see how it works, we need to put the eigenvalues in nonincreasing order. 
The ordering is determined by the ordering of the quantities $e^{\beta E_S}p_{E_S,g}$.
This determines the order of $p(E_S,g)$ (which in general will not be in decreasing order anymore). 
We shall denote such ordered probabilities as $p_i$, 
and the associated energy of the eigenstate as $E_i$.
E.g. $p_1$ is equal to the $p(E_S,g)$ such that $e^{\beta E_S}p(E_S,g)$ is the largest.
Note that for fixed $E_S$ the order is the same as the order of $p({E_S,g})$,
while for different $E_S$ it is altered by the Gibbs factor. We do the same for $\sigma$, 
which results in $q_i$.

\begin{figure}[!ht]
  \centering
  \includegraphics[width=15cm]{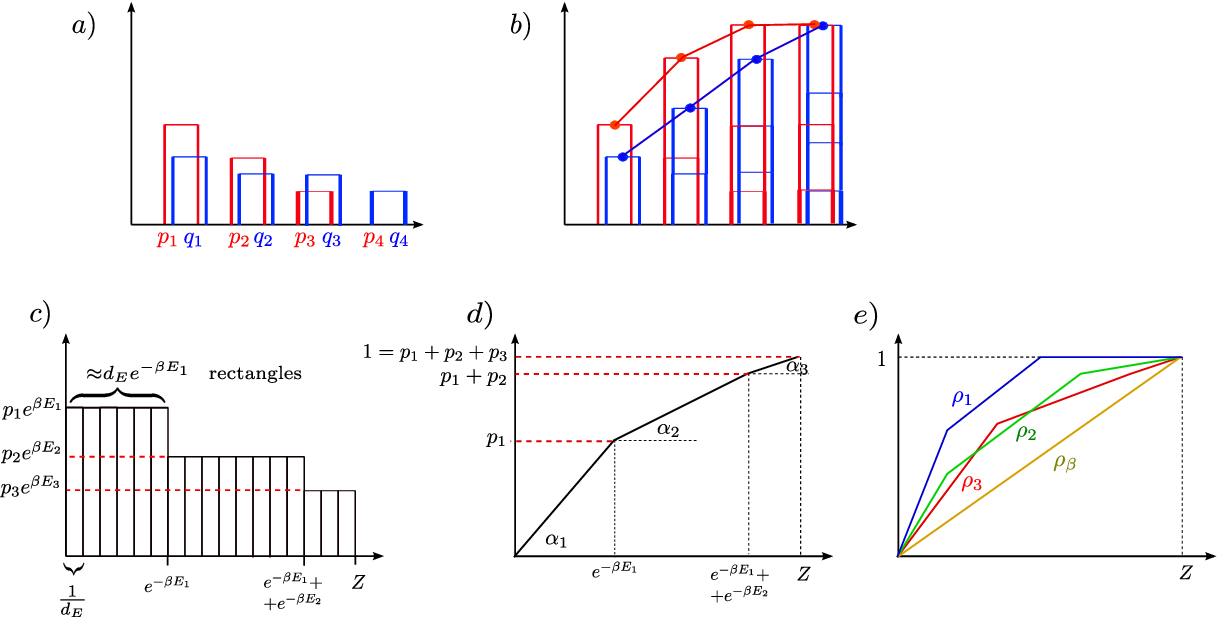}
  \caption{Thermo-majorization. I. Standard majorization: (a) the histograms of probability distributions $\{p_i\}$ and  $\{q_i\}$. (b) The distribution $\{p_i\}$ majorizes distribution $\{q_i\}$
  if for all $l$ $\sum_{i=1}^l p_i\geq \sum_{i=1}^l q_l$. Graphically,this means that  the entire
  plot corresponding to  $\{p_i\}$ is above the plot corresponding to $\{q_i\}$. II. Thermo-majorization.
  (c) Here the histograms consist of groups of numerous columns of the same height. We set their base to $1/d_E$ where $d_E=g_R(E)$, which defines a stair-way looking function defined on interval $[0,Z]$, 
where $Z$ is the partition function. As a result, the plot analogous to that of (b), in the limit of 
large $E$ (which implies $d_E\to \infty$) becomes integral of that function presented on panel (d). 
The angles are given by $\tan(\alpha_i)=p_ie^{\beta E_i}$ hence they are decreasing. (e) checking thermo-majorization conditions amounts to comparing the plots which are piecewise linear functions.
The state $\rho_1$ thermo-majorizes each other state, while the thermal $\tau$ is thermo-majorized 
by all other states. Thus we can transform $\rho_1$ into $\rho_2$, $\rho_3$ and $\tau$ 
and all states can be transformed into $\tau$. On the other hand, $\rho_2$ and $\rho_3$ are incomparable,
hence neither can be transformed into one another by Thermal Operations without an additional work system $\psi_W$.}
\label{fig:thermo-maj-idea}
\end{figure}

The eigenvalues are thus ordered by taking into account Gibbs weights:
\be
\underbrace{\frac{p_1 e^{\beta E_1}}{{d_E}}}_{{\text{multiplicity}\atop \approx d_E e^{-\beta E_1}}}
\geq 
\underbrace{\frac{p_2 e^{\beta E_2}}{{d_E}}}_{{\text{multiplicity}\atop \approx d_E e^{-\beta E_2}}}
\geq \ldots 
\label{eq:beta-order2}
\ee
where $d_E$ is a shorthand for $g_R(E)$.
We shall now ascribe to vector $\{p_i\}$ a function mapping interval $[0,Z]$ into itself. 
On the $y$ axis, we put subsequent sums $\sum_{i=1}^l p_i$, $l=1,\ldots ,d$ where 
$d$ is the number  of all probabilities, 
and on the $x$ axis, we put sums $\sum_{i=1}^l e^{-\beta E_i}$, with the final point being at $x=Z$.
This gives $d+1$ pairs: $(0,0), (p_1, e^{-\beta E_1}), (p_1+p_2, e^{-\beta E_1}+e^{-\beta E_2}), \ldots,
(Z,1)$. We join the points, and it will gives us a graph of a function, $f_p(x)$.
It is easy to see, that in the limit of large $g_R(E)$, the eigenvalues of $\rho$ majorize eigenvalues of $\sigma$  
if and only if $f_p(x)\geq f_q(x)$ for all $x\in[0,Z]$. The described scheme is presented in
\ref{fig:thermo-maj-idea}.

Note that the Gibbs state in this picture is represented by a trivial function $f_\beta(x)=Zx$
hence any state can be transformed into a Gibbs state. Note that one can generalise our new type of majorization,  by replacing the Gibbs state with an arbitrary state, obtaining  an interesting mathematical generalisation of standard majorization.  Likewise, although here the relevant conserved quantity is energy, one can generalise to operations which commute with any conserved quantity.

\section{Transitions involving pure excited states}
\label{sec:pure}

In preparation for deriving the expression for extracting work from a resource, or forming a state from the thermal state
by adding work, we will derive the necessary and sufficient condition for transitions involving a pure energy eigenstate.  In particular, we will derive
the expression for extracting a pure excited state, and the expression for forming a state from a pure excited state.  Then in
\ref{sec:work}, we will use the results in this section to derive our two free energies. 

\subsection*{Distillation: extracting a pure excited state}
\label{sec:dist}
In this section  we derive the condition for when a given mixed state $\rho_S$ with Hamiltonian $H_S$ 
can be transformed into a pure excited state $\psi_W$ - an eigenstate of the 
Hamiltonian $H_{S'}$ with eigenvalue $W$. 
Let us first consider the case where we wish to extract $\psi_W$  with no probability of failure, from
a state diagonal in the energy basis.  We will then extend our result to arbitrary states.

According to Lemma \ref{lem:add_tau} we need  
to  take  an initial state $\rhos\ot \tau_{S'}$ and the final state 
is an arbitrary state of the system $SS'$ of the form  $\sigma_S\ot|\psi^W\>\<\psi^W|_{S'}$. 
Due to Theorem \ref{thm:major}, and Eq. \eqref{eq:g_r_sprim} a transition is possible when the state 
\be
\bigoplus_{E_S}\eta_{E-E_S}^{RS'}\ot P_{E_S} \rhos P_{E_S}
\label{eq:in}
\ee
majorizes 
\be
\bigoplus_{E_S}\eta_{E-E_S-W}^R\ot P_{E_S} \sigma P_{E_S}\ot |\psi_W\>_{S'}\<\psi_W|
\label{eq:out}
\ee
However, since $\sigma$ is arbitrary,  and the target state of $S'$ is pure,
this is equivalent to the condition
\be
\rk_{\rm in}\geq \rk_{\rm out}.
\label{eq:rk}
\ee
where $\rk_{\rm in}$ and $\rk_{\rm out}$ are ranks of the state \eqref{eq:in} and \eqref{eq:out},
respectively.

The rank of the initial state is equal to 
\be
\rk_{\rm in}=\sum_{E_S} g_{RS'}(E-E_S) \rk_{E_S} (\rhos)
\ee
where $\rk_{E_S} (\rho)$ is the rank of $P_{E_S} \rhos P_{E_S}$, 
and as in   Eq. \eqref{eq:g_r_sprim} 
\be
g_{RS'}(E-E_S)=g_R(E-E_S)Z_{S'}.
\label{eq:g_r_sprim2}
\ee
The maximal rank of the target state is given by 
\be
\rk_{\rm out}=\sum_{E_S} g_R(E-E_S-W) g_{S}(E_S)
\label{eq:rk_out}
\ee

Now, using \eqref{eq:g_r_sprim2} and $g_R(E+\Delta E) \approx g_R(E) e^{\beta \Delta E}$ we obtain that 
eq. \ref{eq:rk} implies 
\be
\sum_{E_S}\frac{e^{-\beta(E_S)}}{Z}\rk_{E_S}(\rhos) \leq \sum_{E_{S'}}\frac{e^{-\beta(E_{S'})}}{Z'}\rk_{E_{S'}}(\psi_W)
\ee
which can be written as 
\be
D_{\min}(\rho_S||\tau_S) \geq D_{\min}(\psi^W_{S'}||\tau_{S'})
\label{eq:dmincondition}
\ee
with $D_{\min}(\rho_S||\tau_S):=-\ln\tr{\Pi_{\rho}\gibbs}$.  In general this quantity is the min-relative entropy 
[30,31].

We can now ask about the case when $\rho$ is not diagonal in the energy eigenbasis.  In such a case, we simply replace $\rho$ with 
$\rhodec=\sum_{Eg,g'}\proj{E,g}\rho\proj{E,g'}$ in Equation (\ref{eq:dmincondition}).  
The reason, is that Theorem \ref{thm:major} states necessary and sufficient conditions 
for transforming the diagonal entries of one density matrix into the diagonal entries of another. 
In the case of an initial state with off-diagonal entries, it gives necessary conditions.
However the diagonal entries of a pure excited energy eigenstate determines uniquely that state itself, thus the condition must also
be sufficient.  An alternative argument in terms of commuting of the dephasing operation and thermal operations is given in the \mainsection.


Note that the operation which gets implemented to map one state to another is simply a mapping from eigenstates of the initial state
within each energy block $E$, to mappings of eigenstates of the final state within the same energy block.  However, any such mapping will do, and
there are a huge number of them.  Thus the experimenter does not need to know which unitary she is implementing, provided that it conserves energy.
She thus needs very little control over her systems -- she simply chooses any unitary which maps the macroscopic variables of one state (in this 
case, total energies $(E_R,E_S)$, to macroscopic variables of the final state (in this case, a pure energy eigenstate 
with no degeneracy on some system, and total energy on another $(E_R+E_S-W)$.  The same is true of the formation process described in the next section.

\subsection*{Formation of a resource state from a thermal bath and pure excited state}
\label{sec:form}

Just as one can draw work from a state which is out of equilibrium from the rest of the thermal bath, 
it is also possible to perform the reverse process -- create a state from the thermal bath by adding work.
Here we provide necessary and sufficient conditions for transition from a pure excited state to a given target diagonal state. We will then
use it in \ref{sec:work} to derive the amount of work which is required to create a state.

We thus take the initial state to be of the form 
\be
\rho^{\rm in}=\psi^W_S\ot \tau_{S'}
\ee
and the output state 
\be
\rho^{\rm out} =\rho_{SS'}
\ee
We shall now use Theorem \ref{thm:major}.
To this end we have to check the majorization condition between the following states:
\be
\ideta_{E-W}^{RS'}\ot |\psi^W\>_{S}\<\psi^W|
\label{eq:form_initial}
\ee
and 
\be 
\oplus_{E_{S'}} \ideta_{E-E_{S'}}^{RS}\ot p_{E_{S'}}\rho^{E_{S'}}_{S'}
\ee
where in \eqref{eq:form_initial} we have used Eq. \eqref{eq:g_r_sprim}.

However, the former state has only one eigenvalue $1/g_{RS'}(E-W)$ with multiplicity $g_{RS'}(E-W)$.
Therefore, the majorization condition is that all eigenvalues of the latter state 
are no greater than this eigenvalue. I.e. we need that 
\be
g_{RS'}(E-W)^{-1}\geq g_{RS}(E-E_{S'})^{-1} \lambda^{\max}_{E_{S'}}
\label{eq:majorE}
\ee
holds for all $E_{S'}$, where $\lambda^{\max}_{E_{S'}}$ is the maximal eigenvalue of 
$P_{E_{S'}} \rho_{S'} P_{E_{S'}}$ i.e. it is the maximal eigenvalue of $\rho_{S'}$ 
in the subspace of energy $E_{S'}$. Since the Hamiltonian for $RSS'$ 
is the sum of $H_R$, $H_S$ and $H_{S'}$, we obtain that 
\be
&&g_{RS'}(E-W)=\sum_{E_{S'}} g_R(E-W-E_{S'}) g_S'(E_{S'}) \nonumber \\
&&g_{RS}(E-E_{S'})= \sum_{E_S} g_R(E-E_{S'}-E_S) g_S(E_S) 
\ee
Now we use the fact that $R$ is a heat bath, and we apply our assumption \eqref{eq:g-property}
which says that 
\be
g_R(E-W-E_{S'})\approx g_R(E) e^{-\beta ( W +E_{S'})}  
\ee
and 
\be
g_R(E-E_{S'}-E_S) \simeq g_R(E) e^{-\beta ( E_S +E_{S'})}  
\ee
we can thus rewrite the majorization condition \eqref{eq:majorE} 
as follows
\be
\frac{1}{Z_{S'}}e^{-\beta E_{S'}} \geq \frac{1}{Z_S}e^{-\beta W} \lambda^{\max}_{E_{S'}}
\ee
for all $E_{S'}$.
On the other hand, one can compute that 
\be
&&D_{\max}(\psi^W_S||\tau_S) = Z_S  e^{\beta W}, 
&&D_{\max}(\rho_{S'}||\tau_{S'}) = \max_{E_{S'}}  Z_{S'} e^{\beta E_{S'}} \lambda^{\max}_{E_{S'}}
\ee
where $D_{\max}(\rho||\tau):=\log\min \{\lambda:\rho\leq\lambda\gibbs \}$ is the max-relative entropy [30,31].
Thus, the transition $(\psi^W_S,H_S) \to (\rho_{S'},H_{S'})$ is possible if and only if 
\be
D_{\max}(\psi^W_S||\tau_S) \geq  D_{\max}(\rho_{S'}|\tau_{S'}).
\label{eq:dmaxcondition}
\ee

\section{Extractable work, and work of formation}
\label{sec:work}

We now use the results of 
\ref{sec:pure}
to quantify the amount of work 
that can be drawn from a system in contact with a heat bath of temperature $T$, and the amount of work 
that  is needed to create one. In thermodynamics, both quantities are equal and are given by free energy.
In our case we obtain two free energies, $\fmin$ governing extracting work, 
and the other, $\fmax$, governing creation of the system.  In this section, we derive the expression for $\fmin$ and $\fmax$ in the case where we wish to extract the full amount of work available, or create a total state out of thermal states.
This corresponds to Equations (\ref{eq:ergotropy}) and \eqref{eq:ergotropymax}.
The more general result of Equations \eqref{eq:gmin} and \eqref{eq:gmax} following from thermo-majorization is contained in the \mainsection. We shall first present the so-called "single-shot" results i.e. 
exact transitions between states. Then we will consider "smoothing",
i.e. the processes, where a small probability for failure is allowed. 

\subsection*{Exact transformations}
We propose to define the process of drawing or spending work as raising or lowering the energy level
of an eigenstate of a Hamiltonian $\Hw$ of a system.  This system is used to store the energy provided 
by drawing work. Thus we draw work $\Delta W$
if we transform a state $|E\>$ into $|E'\>$ such that $E'-E=W$, $H_W|E\>=E$ and $H_W|E\>=E$. 
Expending work, would mean the reverse process.  Since our results don't depend on the system used to store work,
we take the most elementary system than can be used, namely a two level system with energy gap $W$.

Thus consider a system $S$ in state $\rhos$. We add a work system with Hamiltonian $\Hw$ 
in a state $|E\>$. Our initial state is thus $\rhos\ot \proj{E}$  and the final state 
$\proj{E'}$. Using the results of 
\ref{sec:pure}
we obtain, that 
$\rhos\ot \proj{E}$ can be transformed into  $\proj{E'}$ if and only if 
\be
D_{\min}(\rhos\ot \proj{E})\geq D_{\min}(\proj{E'}),
\label{eq:drawing-work}
\ee
where we use the shorthand notation 
$D_{\min}(\rho) \equiv D_{\min}(\rho|\tau)$. Since $D_{\min}$ is additive, 
and for energy eigenstates $|E\>$ we have 
\be
D_{\min}(|E\>)= \beta E -\ln Z_W
\ee
where $Z$ is the partition function of the work system, 
we can rewrite \eqref{eq:drawing-work} as 
\be
k T D_{\min}(\rhos)\geq  W
\ee
This allows us to define the free energy $\fmin$ as follows:
\be
\fmin=  F_\beta + kT \dmin
\ee
where $F_\beta$ is the standard free energy of the equilibrium state (we have anyway that for thermal states $\fmin=F_\beta$). 
The work that can be drawn from a 
non-equilibrium state is thus equal to the  the free energy difference $\Delta \fmin$:
\be
W_{\rm dist}(\rho) = \fmin(\rho) - \fmin(\tau)
\ee

Analogously we define work which is needed to create a system, i.e. 
we consider a transition $\proj{E}\to \proj{E'} \ot \rhos$, and in an analogous way obtain
that the minimal work $W=E'-E$ to ensure this transition is given by 
\be
W_{\rm form}(\rho) = \fmax(\rho) - \fmax(\tau)
\ee
where  $\fmax$ is a max-free energy given by 
\be
\fmax=F_\beta + kT \dmax.
\ee
This comes from simply solving Equation \eqref{eq:dmaxcondition} for the value of $W$ required for the transition, to obtain
\begin{align}
W=
kT\inf_{\rho}
\log\min \{\lambda:\rho\leq\lambda\gibbs \}
\end{align}

\subsection*{Transformations allowing failure with a given probability}
\begin{figure}[!ht]
\centering
  \includegraphics[width=6cm]{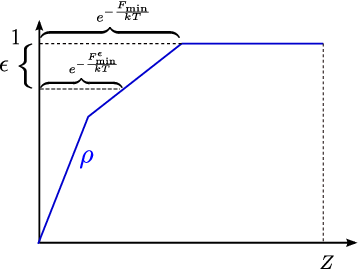}
\caption{Smoothing $\fmin$. Comparison of $\fmin$ and $\ergotropymin$ for given failure probability $\epsilon$.}
\label{fig:smooth-min}
\end{figure}

We now wish to allow some probability of failure [23] -- 
namely, we might not produce $\psi^W_{S'}$ exactly, but rather a state $\rho^W_{\epsilon}$
$\epsilon$-close to $\psi^W$ i.e. such that 
\begin{align}
\<\psi_W |\rho^W_{\epsilon}|\psi_W\> \geq 1- \epsilon.
\label{eq:epsilonclose}
\end{align} 
We now demonstrate the necessary and sufficient condition for this.
By twirling (see \ref{sec:major}) we can assume without loss of generality, 
that the final state is a mixture of $\psi_W$ and some $\rho'$, orthogonal to it (and diagonal 
in the energy eigenbasis):
\be
\rho_W^{\rm out}=(1-\epsilon)|\psi_W\>\<\psi_W| + \epsilon \rho'
\ee 
We shall first consider a protocol which achieves some value of $W$, for given $\epsilon$ and 
for a fixed energy block. We consider an initial state $\tau_R\ot\rho_S\ot |0\>\<0|_W$  and final 
state $\tau_R\ot \sigma_S\ot \rho_W^{\rm out}$. 
Let us also consider a block of fixed energy $E$. In order 
to ensure that the state $\psi_W$ has weight at least $1-\epsilon$ 
we have to map strings of such weight to a subspace of our energy block determined by having 
energy $W$ on system $W$. The size of the space, denoted by $\rk_{\rm out}$ 
is given by \eqref{eq:rk_out}, i.e. 
\be
\rk_{\rm out}=\sum_{E_S} g_R(E-E_S-W) g_{S}(E_S)\approx g_R(E)Z_S e^{-\beta W}
\ee
Thus it is decreasing in $W$. Hence, if we want to have maximal work, then, given $\epsilon$ we have 
to find the smallest group of strings that have total weight no smaller than $1-\epsilon$. 
This group will be mapped onto the above subspace. The remaining ones will be mapped 
onto the strings corresponding to $\rho'$. 
We can thus equally well aim at determining the group of remaining strings, 
which we want to be the most numerous, with the total weight not exceeding $\epsilon$. 
Similarly as in sec. \ref{sec:dist} we find each nonzero eigenvalue $p(E_S,g)$ of $\rho_S$ brings in 
\be
\sum_{E_S}g_R(E-E_S)\approx g_R(E) e^{-\beta E_S}
\ee
strings and  the weight of a single string is $p(E_S,g) e^{\beta E_S} /g_R(E)$. 
Thus, since we want to remove the maximal number of strings, we should start with 
removing ones with the smallest weight. Let us note, that the strings of the smallest weight 
corresponds to the smallest $p(E_i)$ according to the $\beta$-ordering. This leads to the following algorithm: 
let $p(E_i)$ be $\beta$-ordered eigenvalues of $\rho_S$ as in \ref{sec:major} 
(we now use the convention that the index $i$ includes both $E_S$ 
and the degeneracy $g$). Let $l$ be such that 
$\sum_{i=1}^l p(E_i)\leq \epsilon$ and  $\sum_{i=1}^{l+1} p(E_i)\geq \epsilon$.
By Equation \eqref{eq:dmincondition} the possibility of transition is equivalent to 
\be
\sum_{i=1}^l e^{-\beta E_i} + \frac{\epsilon'}{p(E_{l+1})} e^{-\beta E_{l+1}}\leq  Z_S e^{-\beta W}
\ee
where $\epsilon' = \epsilon - \sum_{i=1}^l p(E_i)$. 
This implies that the maximal work, given that we tolerate probability of failure 
$\epsilon$ is given by 
\be
W= \ergotropymin -F(\tau) 
\ee
determined by the following smoothed version of $\fmin$ 
\be
\ergotropymin(\rho) = -kT\ln\sum h(\rhodec,\epsilon,g,E_i) e^{-\beta E_i}
\label{eq:smoothmin}
\ee
where $h(\rhodec,\epsilon,g,E_i)=0$ for $i<l$, $h(\rhodec,\epsilon,g,E_i)=\epsilon'/p(E_i)$ for $i=l$ 
and $h(\rhodec,\epsilon,g,E_i)=1$ for $i>l$, with $\epsilon'$ and $l$ determined using the method described above.
We illustrate this in  \ref{fig:smooth-min}. 
%
The above protocol clearly gives the maximal work for a given energy block. 
However, the maximal work is by definition a monotonic function 
of  $\epsilon$, and the function does not depend on $E$, thus 
the maximal work for a given $\epsilon$ will be the same for the total 
state as for the block.



Now, let us pass to the formation process, provided we are allowed to produce a state $\rho'$ 
which is $\epsilon$ close to the required state $\rho$ in trace norm.
The amount of work needed is given by 
\be
W_\epsilon= \ergotropymax(\rho) - F(\tau)
\ee
where 
\be
\ergotropymax(\rho)=\min_{\rho'}\fmax(\rho')
\ee
where the minimum is taken over all $\rho'$ satisfying $||\rho'-\rho||\leq \epsilon$. 
Here we describe the algorithm to compute $\ergotropymax$ for a given 
state (diagonal in the energy eigenbasis). Namely, 
one considers $\beta$-ordered eigenvalues of the state. In the first stage, 
one subtracts from the $\beta$-largest eigenvalue, and adds to the $\beta$-smallest one,
until either the slope of a line corresponding to the $\beta$-largest on a thermo-majorization diagram 
(see \ref{fig:thermo-maj-idea}d) will become equal to the slope of the second $\beta$-largest,
or the slope of the $\beta$-smallest will become equal to the slope of the second $\beta$-smallest one. 
In the first case, the next step is to subtract from both eigenvalues, again, 
until their slope will become equal to the third $\beta$-largest eigenvalue,
or until the slope of the $\beta$-smallest eigenvalue will become equal to the second $\beta$-smallest one.
In the second case we proceed analogously. We continue, until the total amount of subtracted 
eigenvalue achieves $\epsilon$, or all slopes are equal. Note that the defined algorithm, when optimal ensures that we will not be
left with some remaining part of $\epsilon$ at the end.


\section{Catalytic transitions}
\label{sec:trumping}
So far we have considered the situation 
where the state $\rho$ describes all parts of the system of interest except for the heat bath $\gibbs$.
However, we might want to consider $\rho$ to be a system of interest, interacting with an environment, ancillas, or working body that might be used, 
 but should be returned in it's original state.
Such a process is called catalysis. In [41] we consider this issue in more detail, and here we mention 
how the results are related to the present findings. Let us recall, that the phenomenon of catalysis was 
discovered in entanglement theory, but it also directly translates into the theory of manipulations where purity (or negentropy) is a resource,
where the allowed class of operations (called {\it noisy operations}) consist of (i) adding a system in maximally mixed state,
(ii) arbitrary unitary transformations (iii) tracing out.
Without using ancillas, the transitions are governed by majorization, i.e. 
we can transform probability distributions $p$ into $q$ if and only if $p$ majorizes $q$ (as in Eq. \eqref{eq:major}))
which we denote by $p\succ q$. However, in [42] 
it was shown that some forbidden transitions can be possible, if we can use additional system as a catalyst, i.e. 
we may have $p\not \succ q$ but $p\ot r \succ q \ot r$  for some distribution $r$. One can then ask, what conditions 
should be satisfied by $p$ and $q$ such that an $r$ exists for which $p\ot r \succ q \ot r$. The conditions have been found 
in [47,48] 
and they are called trumping conditions. 
If we are allowed to obtain instead of $q$, an arbitrary good approximation of it, 
then the conditions are the following [41] 
\be 
&& H_\alpha(p) \geq H_\alpha( q), \quad \text{for}\, \alpha \geq 0 \\ \nonumber 
&& H_\alpha(p) \leq H_\alpha(q), \quad \text{for}\,  \alpha < 0
\ee
for $\alpha\in R$, where $H_\alpha$ is the Renyi entropy
\be
H_\alpha(p)= \frac{1}{1-\alpha}\log \sum_i p_i^\alpha
\ee
with $H_1(p)= -\sum_i p_i \log p_i$, $H_0=\log\rank(p)$
where $\rank(p)$ is number of nonzero elements of $p$. Note that the original conditions are in the form of strict inequalities,
however due to strict Schur concavity (convexity) of $H_\alpha$ for $\alpha>0$ ($\alpha<0$), one can change them 
into nonstrict ones.
The above conditions determine possible catalytic transitions with trivial Hamiltonian. 
The conditions with negative $\alpha$ can be removed, if we are allowed to invest an arbitrarily small amount of work,
or, equivalently, to borrow one pure qubit, and return it with arbitrary good fidelity. 

Let us now turn to the case of a nontrivial Hamiltonian, which we consider in this paper. 
In [41] we prove, that 
the necessary and sufficient conditions are expressed in terms of 
Renyi divergences $D_\alpha(\rho|\tau)$ defined (for diagonal states) as 
\be
D_\alpha(\rho\|\sigma)= \frac{1}{\alpha-1} \log \sum_i p_i^\alpha q_i^{1-\alpha}
\ee
for $\alpha>0, \alpha\not=1$ where $p_i$, $q_i$ are the eigenvalues of $\rho$ and  $\sigma$, respectively,
and 
\be
D_\alpha(\rho\|\sigma)= \frac{1}{1-\alpha} \log \sum_i p_i^\alpha q_i^{1-\alpha}
\ee
for $\alpha<0$. For $\alpha=0,1$ we take $D_1(\rho|\sigma) = \sum_i p_i \log (p_i/q_i)$ to be the standard relative entropy
and $D_0(\rho|\sigma)=D_{\min}(\rho|\sigma)$ is the min-relative entropy.
Now, for states  block diagonal in the energy eigenbasis, $\rho$ can be catalytically transformed into 
an arbitrarily good approximation of $\rho'$ by thermal operations, if and only if 
\be
D_\alpha( \rho\|\tau) \geq  D_\alpha(\rho'\|\tau)
\label{eq:D_trump}
\ee
for all $\alpha \in R$. Again, investing an arbitrarily small amount of work, one can remove the conditions with negative $\alpha$.
Alternatively, conditions with negative $\alpha$'s are removed, when, exactly as in the case of the trivial Hamiltonian,
one borrows one qubit in a pure state (with arbitrary Hamiltonian, which can be the trivial one) and returning it 
with arbitrary good fidelity.  The conditions allow for some transitions that are not admitted by thermo-majorization. 
However, they do not affect the
conditions for work of formation and distillation derived in this paper. 
Indeed, for the eigenstates of energy, the divergences with $\alpha\geq 0$ are all equal to the 
standard relative entropy. Moreover the  Renyi divergence is increasing in $\alpha$ for $\alpha\geq 0$, so that,  
in particular, we have $D_{\min}\leq D_\alpha \leq D_{\max}$. Hence, in the case of distillation and formation, the
conditions collapse to a single one: $D_{\min}$  and $D_{\max}$, respectively. Thus the effect of catalysis 
does not affect the thermo-majorization laws in this case. We note further that for $0 < \alpha <2$
a quantum version of $D_\alpha( \rho\|\tau)$ is monotonically decreasing under CP maps [30], and in particular, under
Thermal Operations, and thus for these values of $\alpha$, monotonicity of $D_\alpha( \rho\|\tau)$ provides a necessary condition
for thermodynamic state transitions even for states which are not block diagonal in the energy eigenbasis.

If we allow the catalyst to be returned in a final state only close to its initial state, the crucial point is to choose correctly  the distance measure, since for large systems, the standard distance (trace norm) 
allows the initial and final state of the catalyst to differ arbitrarily with respect to energy or entropy, 
thereby nullifying even the standard Second Law. In the theory of quantum resources  this manifests itself via the phenomenon of embezzling - where one can perform arbitrary transformation, 
while returning a non-exact state with dimension large enough [49]. 
A natural condition that may be imposed, to avoid the phenomenon of embezzling is that 
it should be possible to obtain the required exact output from the approximate one (the one actually obtained) 
by investing some $\epsilon$ amount of work in addition. 
This implies the following (non-symmetric) distance we term the ``trumping distance''
\be
d(\rho\>\rho')=\sup_{\alpha>0} (D_\alpha(\rho\|\tau)-D_\alpha(\rho'\|\tau)).
\ee
In the smoothing procedure one should therefore use the above distance. 
This again does not change the conditions for smoothed work of formation and distillation.
However, one can come up with other criteria, that interpolate between having no Second Law whatsoever,
and the present limitations. The choice of criteria, that best fit the spirit of thermodynamics 
require further investigation.

\section{Characterisation of Thermal Operations}
\label{sec:gibbspreserving}
We have provided an algorithm for deciding whether a state can be transformed into another state,
given by thermo-majorization. However the algorithm does not tell us what kind of 
operations (completely positive maps) we can perform by means of Thermal Operations. 
Below we shall show, that all possible processes are precisely those that preserve the Gibbs state.
This was shown for a special case of tensor product of i.i.d states [25].  
Here, we show that it is true for a generic heat bath that satisfies our assumptions. 
This implies that if we have reversibility of state transformations (as is the case when we have many copies of a state [21], 
then the unique measure which determines whether a transformation is possible, is given by the relative entropy distance to the Gibbs state [28]. This quantity is the difference between free energy of a state of interest and that of the Gibbs state [19]. However, here, we do not have reversibility, thus there are at least two inequivalent functions which are non increasing under thermal operations  ($\fmin$ and $\fmax$).

We start with a state $\tau_R\ot\rho_S$ and write 
\be
\tau_R\ot\rho_S \approx \sum_{E\in \ecal} p_E \rho_{RS}^E 
\ee
with
\be
\rho_{RS}^E= \frac{1}{P_E}P_E\tau_R\ot\rho_S P_E
\ee
where $\ecal$ consists of very large energies in comparison with system energies,
and $p_E=\tr(\tau_R\ot\rho_S P_E)$. 
We shall now fix one energy block, and show, that even when restricting just to 
permutations of basis vectors within the block (being products of eigenstates of $\tau_R$ to eigenvalues $E_R$  and eigenstates of $\rho_S$ to eigenvalues $E_S$, such that $E_R+E_S=E$) we 
can perform arbitrary operation on system $S$ which preserve the Gibbs state. 
Then, we will argue that the operation on the system $S$ can be made the same 
for each energy block (for $E\in \ecal$).

\begin{figure}[!ht]
  \centering
  \includegraphics[width=10cm]{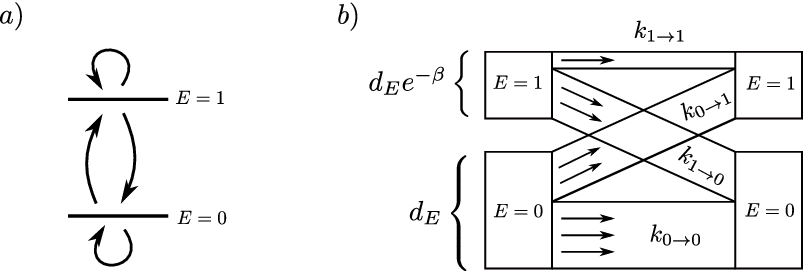}
  \caption{Transitions for two level system with energy levels $E=0$ and $E=1$. a) There are four possible transitions b) the number of states of bath corresponding to each level is proportional 
  to the Boltzmann factor.}
\label{fig:processes}
\end{figure}

To prove the first claim, for simplicity, 
let us assume that the Hamiltonian $H_S$ is nondegenerate (extension to the degenerate case is immediate). 
As follows from Theorem \ref{thm:major}, in such a fixed subspace, the eigenvalues of 
our state form groups labelled by energy $E_S$. Within each group, we have 
$g_R(E) e^{-\beta E_S}$ eigenvalues all equal to $\frac{p(E_S)}{g_R(E)e^{-\beta E_S}}$. 
Permutations of basis vectors result in transferring some subsets of a given group to other groups. 
Let us then use indices $i$ in place of $E_S$, so that $p_{E_S}\to p_i$ and  $g_R(E)e^{-\beta E_S}\to d_i$. 
We shall denote by $k_{i\to j}$ the "transition current" i.e. the number of eigenstates that have been moved from the $i$-th group to the $j-th$ group.  
Clearly $k_{i\to j}$ satisfy
\be
\sum_j k_{i\to j} =d_i\\
\label{eq:k1}
\quad \sum_i k_{i\to j}=d_j\\
\label{eq:k2}
\ee
The transition "currents" are illustrated on \ref{fig:processes}.
After an operation given by some fixed set of $k_{i\to j}$ satisfying the above transitions, we obtain 
a new state, whose probabilities $q_i$ are given by 
\be
q_j=\sum_i k_{i\to j} \frac{p_i}{d_i} 
\ee
Thus, we can define transition probabilities $\pij$ as 
\be
\pij=\frac{k_{i\to j}}{d_i}
\ee 
Then the condition \eqref{eq:k1} means that the $\pij$ ensure normalisation, so 
that the only constraint on possible process is \eqref{eq:k2}. However, since 
$\frac{d_i}{d_j}=\frac{e^{\beta E_j}}{e^{\beta E_j}}$, the latter condition 
means simply that the Gibbs state is preserved. This ends the proof,
that for fixed $E$ we can perform all Gibbs preserving operations. 
Finally, given arbitrary Gibbs preserving transformation on $S$ 
we perform for every total energy block permutation
that results in this transformation. In this way the needed transformation is 
performed on the initial state of system $S$. Of course, Thermal Operations obviously 
do preserve the Gibbs state, hence we obtain, that Thermal Operations are arbitrary operations 
that preserve the Gibbs state.   Furthermore, although a single copy of another state (such as that of Equation \ref{eq:puregibbs})
may also be preserved under Thermal Operations, if we allow many copies of some state, then only the Gibbs state is preserved.
This follows immediately from the fact that the relative entropy distance to the Gibbs state is the unique monotone in the case
of many copies of $\rho$ [21].

\begin{figure}[!ht]
  \centering
  \includegraphics[width=12cm]{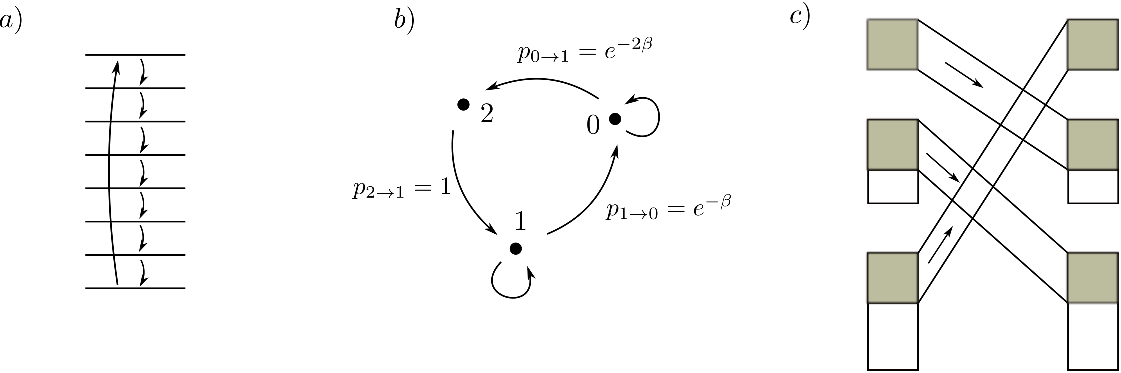}
  \caption{Quasi-cycles. a) In the quasi cycle  for each level there is only one transition to a different level b) a quasi-cycle for 
  a three level system. The points represent levels with energies $0,1$ and $2$; probability $p_{2\to2}$
  of staying in level $2$ vanishes. c) Currents for a three-level quasi-cycle: the shaded microstates 
  are subjected to a cycle, while the ones not shaded are left untouched.}
\label{fig:quasi-cycles}
\end{figure}

Let us discuss this result in the context of the detailed balance condition. The latter 
is the property that $\frac{p_{i\to j}}{p_{j\to i}}=e^{-\beta (E_j-E_i)}$.
As we will  see, Thermal Operations need not satisfy detailed balance; they should merely
preserve the Gibbs state as a whole. To provide an example, let us distinguish a class of Gibbs-preserving processes called quasi-cycles: 
we put the energy levels on a circle, and from one level, one can go only to the next neighbouring level, as in \ref{fig:quasi-cycles}.

The simplest description of a quasi-cycle is in terms of 
quantities $k_{i\to j}$. Namely, we choose an order of levels, put them on a circle, fix 
a direction,  and  the process is to take all states from the group 
of states with the largest energy $E_S$, and shift them to the states with the energy level in the chosen direction. 
I.e. the process is determined by $p_{i\to i+1}=e^{-\beta (E_{\max}-E_i)}$,
where $E_{\max}$ is the maximal energy, and $E_i$ is the energy of the $i$-th level. 

For two level systems, the class of Gibbs preserving operations is the same as the class 
of operations satisfying the detailed balance condition,
and all possible processes are parametrised by a single number $r\in[0,1]$, which is the probability of 
mixing two basic processes: the identity operation, and the two-level quasi-cycle.
For three level systems, there are processes that preserve the Gibbs state, 
but do not satisfy detailed balance, an example being the three-level quasicycle. It turns out  that the class of Gibbs preserving 
maps is strictly more powerful that the class of detailed-balance maps. An example is the transition 
between $(0,\frac12,\frac12)$ and 
$(\frac12e^{-2\beta}, \frac12(1- e^-\beta),\frac12(e^{-\beta}-e^{-2\beta}+1))$ 
with the energy levels given by $(0,1,2)$. It turns out that the only Gibbs preserving operation that can transform  the first state into the second one is the quasi-cycle  $0\to1\to2$.
This means that such a transition is impossible by means of weak coupling with the heat-bath [50],
as at weak coupling the detailed balance condition is satisfied.

\section{Small engines undergoing a Carnot cycle}
\label{sec:carnot}

\begin{figure}[!ht]
  \centering
  \includegraphics[width=8cm]{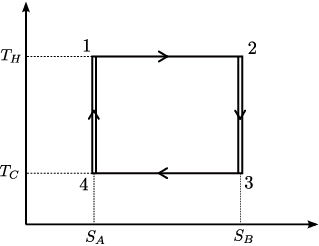}
\caption{Microscopic Carnot cycle. The double line for the isentropic processes illustrates different paths 
depending on whether the system is in the excited or ground state.}
\label{fig:carnot}
\end{figure}

Here, we analyse a microscopic engine undergoing a Carnot cycle, between two heat reservoirs at temperatures $T_H$ and $T_C$ 
-- see \ref{fig:carnot}.  
Since here, we are interested in work extraction from single systems, we are interested in 
how the engine behaves over a single cycle rather than its long term running efficiency. At stage $1$ the working body 
 of the
engine is in a thermal state, and is in contact with a thermal
reservoir at temperature $T$.  The Hamiltonian of the engine we call $H_1$ and we take the working body to be as small as possible - a qubit.  
If we were to then extract maximal work from the working body, we would do so at the rate given by 
Equation \eqref{eq:ergotropy}.  However, the density matrix of the state of the  working body, being thermal, has full rank and so,
no work can be extracted without some probability of failure at least as large as it's smallest eigenvalue, $e^{-\beta E_{\max}}/Z$.  However, there is
a way around this limitation -- we can add a second system which is not in contact with the heat bath, and which
acts as a switch qubit, as we did in Equation \eqref{eq:changingham}, and 
effectively change the Hamiltonian that acts on the working body from $H_1$ to $H_2$.
The amount of work that can then be extracted in the first isothermal stage is then $W_{12}=-kT_H\ln{Z_1/Z_2}$ as given by Equation 
\eqref{eq:thermaltransitions}, and with $Z_i$ being the partition function corresponding to $H_i$.  The work gained is stored in a work system
$W$ which is initially in the ground state.  We can thus perform the isothermal
process reversibly and ideally, provided we add an extra system to simulate the changing Hamiltonian, and don't extract the maximal amount
of work from the engine.

Next we remove the working body from the thermal reservoir, and undergo the first adiabatic process.  This will involve changing the Hamiltonian
acting on the qubit slowly, so that the populations of the ground and excited state doesn't change, and transferring the gained
energy to our work system $W$.  This requires that the switching system have a third state.  At the end of this process (stage 3),
the working body needs to be in a thermal state at temperature $T_C$ in order for the third stage to proceed reversibly. We thus need
\begin{align}
e^{-\beta_HE_2^e}/Z_2&=e^{-\beta_CE_3^e}/Z_3\nonumber\\
e^{-\beta_HE_2^g}/Z_2&=e^{-\beta_CE_3^g}/Z_3
\label{eq:matching23}
\end{align}
where $E_i^g$ and  $E_i^e$ are the ground and excited energy levels of $H_i$.

Here we have a problem: we need to conserve energy, but the qubit
of our working body could either be in the ground state or the excited state.  To extract a deterministic amount of work, when we don't
know whether the working body is in the ground or excited state,  we would need
that the change in energy of the excited state, is the same as the change in energy of the ground state.  i.e. that 
$W_{23}=E_2^e-E_3^e=E_2^g-E_3^g$.  However, it is impossible to satisfy this condition, and those of Equations \eqref{eq:matching23}.
Either we need to return the qubit to the cold bath in a non-equilibrium state, resulting in inefficiencies, or we need to 
extract a different amount of work ($W_{23}^e=E_2^e-E_3^e$ and $W_{23}^g=E_2^g-E_3^g$), depending on whether the working body is in the ground or excited state.
We  therefore  produce 
some  additional entropy in the work system.  In order to have the amount of entropy production 
be small compared with the 
amount of work extracted, one would need to have a larger heat engine -- for example, if we have $n$ two level systems, we find the system
has a higher probability to be around the average energy as $n$ increases, compared with being in the ground state [21], 
allowing us to conserve energy and more deterministically extract work.

Next, we perform the second isothermal stage, this time at temperature $T_C$.  This time, the work is added to the system
and is thus negative $W_{34}=-kT_C\ln{Z_3/Z_4}$ as given by Equation \eqref{eq:thermaltransitions}.  Then we perform the final 
adiabatic process, and again, we have the same problem as before, extracting different amounts of work
$W_{41}^e=E_4^e-E_1^e$ and $W_{41}^g=E_4^g-E_1^g$ depending on whether the engine qubit is in the ground or excited state. 
As before, we also want, that at the end of the final stage, the working body qubit is in equilibrium with $T_H$
\begin{align}
e^{-\beta_C E_4^e}/Z_4&=e^{-\beta_HE_1^e }/Z_1 \nonumber\\
e^{-\beta_CE_4^g}/Z_4&=e^{-\beta_H E_1^g}/Z_1
\label{eq:lipa}
\end{align}
Solving Equations (S72)-(S73) 
we obtain \ref{table:Carnot_1}.
\begin{table}[h]
\begin{tabular}{c|c|c|c|c}
stage &1&2&3&4\\
\hline
$E^e$ & $E$ &$E_2^e= \frac{T_H}{T_C} E_3^e -\delta$ & $E_3^e$ & $E_4^e =\frac{T_C}{T_H} E+\delta'$\\[1mm]
\hline
$E^g$ & $0$ & $E_2^g = \frac{T_H}{T_C}-\delta$ & $E_3^g$ & $E_4^g =\delta'$ \\[1mm]
\hline
$Z$ & $Z_1=1+e^{-\beta_H E}$ & $Z_3 e^{\beta_H \delta }$& $Z_3= e^{-\beta_C E_3^g}+e^{-\beta_C E_3}$ 
& $Z_1 e^{-\beta_H E}$\\[1mm]
\hline 
$p_e$ & $\frac{1}{Z_1}{e^{-\beta_H E}}$ &$\frac{1}{Z_3}{e^{-\beta_C E_3}}$ & $\frac{1}{Z_3}{e^{-\beta_C E_3}}$
&$\frac{1}{Z_1}{e^{-\beta_H E}}$ 
\end{tabular}
\caption{Four stages of the Carnot cycle. $E^e$ and $E^g$ denote the excited and ground states of subsystem $S$ of the working body. 
The probability of being in the excited state is denoted by $p_e$.}
\label{table:Carnot_1}
\end{table}
At the end of a cycle, there are four possible amounts of work extracted, depending on whether the working body qubit was in the 
ground or excited state at stage $2$ and $4$.  We label probabilities of those for events as $p_{ee},p_{eg},p_{ge},p_{gg}$.  Regardless of
what these probabilities are, it's not hard to see that we obtain an amount of work that on average is equal to the amount
obtained by an ideal engine
\begin{align}
W_{avg}&=-kT_H\ln{Z_1/Z_2}-(\frac{T_H}{T_C}-1)\langle E_3\rangle -\delta-kT_C\ln{Z_3/Z_4}-(\frac{T_C}{T_H}-1)\langle E_1\rangle+\delta'\nonumber\\
&=(T_H-T_C)(S_B-S_A)
\end{align}
However, in any single run of the cycle, we will see that there are three possible amounts of work which can be extracted, as 
depicted in \ref{table:Carnot_2}.  
\begin{table}[h]
\begin{tabular}{c|c|c|c|c}
state &ee&eg&ge&gg\\
\hline
$\Delta W$ & $(\frac{T_H}{T_C}-1)E_3^e+(\frac{T_C}{T_H}-1)E$ & $(\frac{T_H}{T_C}-1)E_3^e$ & $(\frac{T_H}{T_C}-1)E_3^g+(\frac{T_C}{T_H}-1)E$ & $(\frac{T_H}{T_C}-1)E^g_3$\\[1mm]
\hline
$\frac{e^{-\beta_H E}}{Z_1}\leq \frac{e^{-\beta_C E_3}}{Z_3}$ & $p_{ee}=\frac{e^{-B_H E}}{Z_1}$ & $p_{eg}=\frac{e^{-\beta_C E_3}}{Z_3}-\frac{e^{-\beta_H E}}{Z_1}$ & $p_{ge}=0$ & $p_{gg}=\frac{e^{-\beta E_3^g}}{Z_3}$ \\[1mm]
\hline
$\frac{e^{-\beta_H E}}{Z_1}\geq \frac{e^{-\beta_C E_3}}{Z_3}$ & $p_{ee}=\frac{e^{-B_CE_3}}{Z_3}$ & $p_{eg}=0$ & $p_{ge}= \frac{e^{-\beta_H E}}{Z_1}-\frac{e^{-\beta_C E_3}}{Z_3}$ & $p_{gg}=\frac{1}{Z_1}$ \\[1mm]
\hline 
\end{tabular}
\caption[l]{Work extracted in single Carnot cycle. Amount of work extracted in any single cycle depends on the state of the working body. There are four possible amounts of work extracted, depending on whether the working body qubit was in the 
ground or excited state at stage $2$ and $4$; we denote those events by ee, eg, ge and gg.  Here, we only write
the difference in the amount of work extracted compared with a basic amount of $W_o=-kT_H\ln{Z_1/Z_3}-kT_C\ln{Z_3/Z_1}$.
There are two cases, one when $e^{-\beta_H E}/Z_1\leq e^{-\beta_C E_3}/Z_3$ and the other when $e^{-\beta_H E}/Z_1\geq e^{-\beta_C E_3}/Z_3$.
In the first case, the entropy produced in a cycle is 
$S(W)=e^{-\beta_HE}/Z_1(\beta_H E +\ln{Z_1})
+e^{-\beta_CE_3^g}/Z_3(\beta_C E_3^g+ \ln{Z_3})
-( e^{-\beta_C E_3}/Z_3-e^{-\beta_H E}/Z_1 )
\ln{(e^{-\beta_C E_3}/Z_3-e^{-\beta_H E}/Z_1)}
$
and in the second case,
$S(W)=\ln{Z_1}/Z_1+e^{-\beta_CE_3}/Z_3(\beta_C E_3+ \ln{Z_3})
+( e^{-\beta_C E_3}/Z_3-e^{-\beta_H E}/Z_1 )
\ln{(e^{-\beta_H E}/Z_1-e^{-\beta_C E_3}/Z_3)}$.  The probabilities are computed under a process which minimises the probability that a transition
between ground to excited or visa versa is undergone, which can be achieved under Thermal Operations.}
\label{table:Carnot_2}
\end{table}

The work storage system thus needs to be at least a 
four level system (initially in state $0$), and the engine is not ideal -- 
entropy is pumped into the work storage system.  Alternatively, one could remove the entropy by letting the higher energy states of the work
system decay into the lowest level excited state.

Note that if we are interested in having the working body of our Carnot engine to be as small as possible, we can take 
\begin{align}
H_2=H_4
\end{align}
so that the switching system need only
be of dimension $3$.  
i.e. $2\times3$ not including the work storage system of dimension $4$ per cycle.  
To approach the Carnot efficiency in a single cycle, might will need even more qubits.  
It would be interesting to compare our small Carnot type engine to the small 
versions [9,51,52] 
of the Brownian ratchet [53,54] 
when operating only a few cycles.

We thus see that small engines pump additional entropy into the system.  In macroscopic systems, or the case of $\rho^{\otimes n}$ discussed in [21],
the amount of entropy produced can be made arbitrarily small.  Note that what is important is not that the spread in energy
be small compared to the average energy (indeed, the Gibbs state $\gibbs^{\otimes n}$ has this property for $n$ large enough).  Rather, we want
that for a fixed average energy, the entropy of the work system needs to be arbitrarily small compared to the entropy of the corresponding heat
bath at that average energy. However,
here, this is not the case.  If we wish to draw a deterministic amount of work, we sometimes fail, and we thus have a failure probability
in micro-engines.  Or rather than allowing a probability of failure, we could run the engine over many cycles, and then remove the additional
entropy that was produced by acting collectively on the many work storage systems.  However, here we are interested in the work produced by single
systems, or during a single cycle, and thus, must be content to tolerate entropy production, an amount of work lower than the optimal amount, 
or a probability of failure.





\end{document}